\documentclass[aps,pre,twocolumn,showpacs]{revtex4}

\usepackage{amsmath}
\usepackage{graphicx}
\usepackage{stackengine}
\usepackage{multirow}
\usepackage[caption=false]{subfig}

\begin{document}
\title{Relaxation dynamics of vortex lines in disordered type-II 
       superconductors following magnetic field and temperature quenches}

\author{Hiba Assi$^{1}$ \email{hiba.assi@vt.edu}}
\author{Harshwardhan Chaturvedi$^{1}$} 
\author{Ulrich Dobramysl$^{2}$} 
\author{Michel Pleimling$^{1}$} 
\author{Uwe C. T\"auber$^{1}$ \email{tauber@vt.edu}}

\affiliation{$^{1}$Department of Physics (MC 0435), 850 West Campus Drive, 
	Virginia Tech, Blacksburg, Virginia 24061}
\affiliation{$^{2}$Mathematical Institute, University of Oxford, 
        Andrew Wiles Building, Radcliffe Observatory Quarter, Woodstock Road, 
        Oxford OX2 6GG, U.K.}

% submitted to Phys. Rev. E (2015); arXiv:1505.06240

\date{\today}

\begin{abstract}
We study the effects of rapid temperature and magnetic field changes on the 
non-equilibrium relaxation dynamics of magnetic vortex lines in disordered 
type-II superconductors by employing an elastic line model and performing 
Langevin molecular dynamics simulations. 
In a previously equilibrated system, either the temperature is suddenly 
changed, or the magnetic field is instantaneously altered which is reflected 
in adding or removing flux lines to or from the system. 
The subsequent aging properties are investigated in samples with either 
randomly distributed point-like or extended columnar defects, which allows to 
distinguish the complex relaxation features that result from either type of 
pinning centers. 
One-time observables such as the radius of gyration and the fraction of pinned 
line elements are employed to characterize steady-state properties, and 
two-time correlation functions such as the vortex line height autocorrelations 
and their mean-square displacement are analyzed to study the non-linear 
stochastic relaxation dynamics in the aging regime.
\end{abstract}


% 05.40.-a - Fluctuation phenomena, random processes, noise, and Brownian motion
% 74.25.Wx - Vortex pinning
% 74.25.Uv - Vortex phases
% 74.40.-n - Fluctuation phenomena
% 74.62.En - Effects of disorder
 
\maketitle

\section{Introduction}
\label{sec:introduction}

Type-II superconductors have been the focus of many theoretical and 
experimental investigations since their discovery. 
From the fundamental physics perspective, these materials represent a class of 
systems with many competing energy scales and thus a remarkably rich variety of
thermodynamic phases, as well as intriguingly complex transport 
properties~\cite{Blatter1994}. 
Applied science and engineering research on type-II superconductors has focused
primarily on their optimization for technological applications, especially in
magnetic fields. 
The most prominent desirable property of superconductors is of course the 
absence of current flow dissipation in these materials. 
However in the mixed phase of type-II superconductors, magnetic flux enters the
sample in the form of quantized vortex lines, and external driving currents 
generate a Lorentz force that in turn causes these vortices to move, creating 
Ohmic dissipation in the system. 
To prevent flux flow and thus restore dissipation-free transport, effective 
pinning mechanisms of flux lines to material defects are 
required~\cite{Blatter1994}. 

Different types of defects are utilized as pinning centers for magnetic 
vortices, predominantly uncorrelated point-like disorder and extended linear 
(columnar) defects. 
Point defects are naturally occurring, e.g., in ceramic high-$T_c$ materials 
in the form of oxygen vacancies, but can also be artificially introduced, for 
instance by electron irradiation~\cite{Kwok1994}. 
The presence of weak point pinning centers destroys the long-range crystalline 
order of the low-tempera\-ture Abrikosov flux line lattice in the disorder-free 
system, to form either a genuine disordered vortex glass 
phase~\cite{Fisher1989,Feigelman1989,Nattermann1990,Fisher1991,Kwok1992} or
a Bragg glass state that is characterized by quasi long-range positional 
order~\cite{Giamarchi1994,Giamarchi1995,Kierfeld1997,Fisher1997,Giamarchi1997,Nattermann2000}. 
The thermally induced first-order melting transition of the vortex lattice at
elevated temperatures~\cite{Nelson1988,Nelson1989,Nelson1989a} is thereby 
replaced by a disor\-der-driven continuous phase transition between frustrated
(`glassy') low-temperature states and a fluctuating flux liquid phase. 

Columnar defects too may be naturally present in the sample as line 
dislocations, and can in addition be artificially introduced by growing them 
using, for instance, MgO nanorods~\cite{Yang1996} or through irradiating the 
sample with high-energy ions such as Sn, Pb, or Iodine. 
In contrast with point-like disorder, the presence of correlated columnar 
defects results in the emergence of a novel low-tempe\-rature thermodynamic 
state distinct from the vortex glass phase, namely the strongly pinned Bose 
glass~\cite{Kwok1992,Nelson1992,Lyuksyutov1992,Nelson1993,Fisher-Weichman1989}. 
Since vortex lines become then localized along the entire length of these 
linear pinning centers, the sample's tilt modulus diverges (transverse Meissner
effect) \cite{Nelson1993,Tauber1997}. 
Indeed, columnar defects have experimentally proven more efficient at pinning 
than uncorrelated point-like disorder due to their extended 
nature~\cite{Civale1991}. 

We emphasize that in the vortex glass phase in the presence of point-like 
disorder, the low-energy flux line contours tend to roughen as the vortices 
minimize their free energy in the disordered pinning landscape, while in the 
Bose glass phase in the presence of columnar defects that are oriented along 
the magnetic field direction the bound vortex lines are effectively straight. 
Therefore, these two disorder-dominated glassy phases in fact display very 
different lateral and transverse vortex line fluctuations. 
One consequently expects samples with point disorder and columnar defects to 
yield profoundly distinct non-equili\-brium relaxation properties. 

Some materials undergo slow relaxation processes to eventually reach thermal 
equilibrium, which affects their measured dynamical properties and renders 
these effectively time-dependent; this is referred to as physical 
`aging'~\cite{Struik1978,Henkel2010}. 
Many glassy systems have been observed to undergo physical 
aging~\cite{Henkel2007} owing to the very slow kinetics between a large number 
of energetically close metastable states typically emerging in frustrated 
environments.  
In a superconducting 2H-NbSe$_2$ sample, Du \textit{et al.} discovered that 
the voltage response to an applied current pulse depended on the pulse 
duration~\cite{Du2007}, which constitutes clear evidence of physical aging in 
disordered vortex matter. 
Papadopoulou \textit{et al.} observed aging effects in the measurement of the 
zero-field cooled magnetization in the compound
Bi$_2$Sr$_2$CaCu$_2$O$_8$~\cite{Papadopoulou1999}. 

On the numerical front, Bustingorry, Cugliandolo, and Dom\'inguez pioneered 
Lan\-gevin molecular dynamics simulations of a three-dimensio\-nal model of 
vortex matter to identify physical aging features in certain two-time 
correlation functions~\cite{Bustingorry2006,Bustingorry2007}. 
Langevin molecular dynamics simulations were also performed by Cao 
\textit{et al.} to study the dynamical scaling properties of a 
three-dimensional vortex line model subject to an ac drive near the critical
depinning threshold~\cite{Cao2012}. 
Pleimling and T\"auber then employed an elastic line model and Monte Carlo 
simulations to investigate the non-equilibrium relaxation properties of 
magnetic vortices in type-II superconductors starting from somewhat artificial
initial configurations where straight flux lines were placed at random 
positions in the sample~\cite{Pleimling2011}. 
The resulting complex aging features (with identical initial conditions and
parameter values) were subsequently confirmed in a very different microscopic 
representation of the non-equilibrium vortex kinetics through Langevin 
molecular dynamics~\cite{Dobramysl2013}. 
A systematic investigation of the rich non-equilibrium relaxation dynamics and 
aging features of complex systems such as disordered vortex matter with various 
competing energy and time scales is of fundamental interest. 
Specifically, it is crucial to assess the relevance of distinct initial 
configurations and their effects on the subsequent slow relaxation kinetics.

In this present work, we investigate in detail the aging relaxation dynamics 
of vortex matter in disordered type-II superconductors starting from initial
conditions that more closely resemble experimentally realizable situations, 
where we analyze the effects of sudden changes in either the system's ambient 
temperature or magnetic field. 
This allows us to both make closer contact to experiment, and to explicitly 
test the sensitivity of the non-equilibrium aging kinetics to the selection of
initial states.
As in our earlier studies, we can switch off the repulsive vortex interactions,
the pinning potentials, or both in our computer simulations, and thereby 
identify the physical mechanisms for the observed complex features in the 
vortex relaxation dynamics. 
Again, crucial differences in the fluctuation spectrum for vortex matter in 
the presence of point-like or extended defects become apparent.
In addition, separately tracking the behavior of the originally present, 
already relaxed flux lines and the newly added vortices helps our analysis of
the simulations with sudden magnetic field increases.

One of the principal goals of this investigation is to identify observable 
correlations that allow a dynamical characterization of material samples. 
Experimental methods that could be used to extract the relevant quantities 
include the following techniques: 
Vortices in type-II superconductors were imaged by Vasyukov \textit{et al.} 
using nano-SQUIDs in scanning probe microscopy~\cite{Zeldov2013}, with 
recorded magnetic fields down to $50\,$nT. 
Furthermore, Auslaender \textit{et al.} used the tip of a magnetic force 
microscope to drag the end of a single vortex across the surface of a YBCO 
sample, directly measuring its interaction with a local disorder potential, in
order to study the dynamics of vortices and their (de-)pinning processes from 
different defects~\cite{Auslaender2009}. 
Flux creep parameters in YBCO samples were experimentally investigated by 
Abulafia \textit{et al.} by utilizing an array of microscopic Hall 
sensors~\cite{Abulafia1997}. 
Structural properties of vortices in superconductors can be imaged through
small-angle neutron scattering~\cite{Eskildsen2011}, a technique that 
directly measures the Fourier transform of the flux lines' height-height
correlation function, and thus accesses their lateral fluctuations reflected
in the observables introduced in Section~\ref{sec:measured-quantities}. 
These methods could perhaps be utilized to probe the relaxation processes of 
magnetic vortex lines in a disordered medium following quenches in temperature 
or magnetic field, and to distinguish the effects of different types of 
disorder on the dynamics of this complex system. 
The initial configuration of randomly placed straight vortices that were 
implemented in earlier numerical studies cannot be realized in experiments. 
Instead, we investigate the effects of abrupt changes of external experimental 
control parameters, namely the system's ambient temperature and magnetic field.

This paper is organized as follows: 
In the next section, we describe the elastic line model and the Langevin 
molecular dynamics simulation algorithm that we employ in our study, and also
specify the material parameters that we use in our computational model. 
Then, we specify the single-time quantities and two-time correlation functions
that are measured in our simulations to analyze the non-equilibrium and aging 
properties of our vortex matter models.
The last part of this section is devoted to a discussion of finite-size 
effects and a verification that our numerical results are not specific to the
system size we choose. 
Section~\ref{sec:temp} discusses the effects of temperature quenches on the 
relaxation properties in systems of interacting flux lines subject to either
point-like or columnar disorder. 
In Section~\ref{sec:mag}, we present the relaxation dynamics in the aging
regime following magnetic field quenches. 
We compare the effects of sudden increases or decreases of the vortex density
to our earlier studies where the magnetic field remained fixed. 
We systematically disentangle the contributions due to repulsive vortex
interactions and pinning to disorder.
We find the non-equilibrium relaxation features of vortex matter to strongly  
depend on the type of disorder present in the system. 
We finally summarize our work in the concluding Section~\ref{sec:conclusion}.

\section{Model Description and Simulation Protocol}
\label{sec:model-simul-prot}

\subsection{Elastic Line Description}
\label{sec:effect-model-hamilt}

In the following numerical study, we consider systems composed of $N$ vortex
lines in the extreme London limit, where the London penetration depth 
$\lambda$ is much larger than the coherence length $\xi$, which allows us to 
employ a fully three-dimensional elastic line 
description~\cite{Nelson1993,Das2003}. 
The vortices are described through their trajectories $\vec{r}_i(z)$, 
{\em i.e.}, their two-dimensional $ab$ plane position vectors as function of 
the $z$ coordinate along the direction of the applied external magnetic field. 
The system's Hamiltonian can then be written as a functional of the $N$ 
interacting trajectories $\vec{r}_i(z)$:
\begin{equation}
  \label{eq:hamiltonian}
  \begin{split}
    H\left[\vec{r}_i\right]=&\sum_{i=1}^N\int_0^Ldz\Biggl[
   \frac{\tilde{\epsilon}_1}{2}\left|\frac{d\vec{r}_i(z)}{dz}\right|^2
   + U_D(\vec{r}_i(z),z)\\
   &+ \frac{1}{2}\sum_{j\ne i}^NV(|\vec{r}_i(z)-\vec{r}_j(z)|) \Biggr] .
  \end{split}
\end{equation}

This effective free energy consists of three competing terms: the elastic 
vortex line tension, an attractive potential describing the $N_D$ pinning 
centers, and the repulsive interactions between different vortices. 
The elastic line stiffness or local tilt modulus is given by 
$\tilde{\epsilon}_1 \approx \Gamma^{-2}\epsilon_0\ln(\lambda_{ab}/\xi_{ab})$, 
where $\Gamma^{-1} = M_{ab}/M_c$ is the anisotropy parameter (effective mass 
ratio). 
$\lambda_{ab}$ and $\xi_{ab}$ respectively denote the London penetration 
depth and the coherence length in the $ab$ crystallographic plane. 
Moreover, the in-plane vortex repulsive interactions are given by 
$V(r) = 2\epsilon_0 K_0(r/\lambda_{ab})$, with $K_0$ denoting the 
zeroth-order modified Bessel function, which essentially represents a 
logarithmic repulsion that is exponentially screened at the scale $\lambda$. 
We set a cut-off for the interactions in our simulations at $5\lambda_{ab}$ 
to avoid artifacts due to the periodic boundary conditions. 

We discretize the Hamiltonian \eqref{eq:hamiltonian} in the $z$ direction, 
and model $N_D$ point-like pinning centers as spatially randomly distributed 
smooth potential wells of the form
\begin{equation}
  \label{eq:pinningwells}
  \begin{split}
    U_D(\vec{r},z) \!=\!-\!\! \sum_{\alpha=1}^{N_D}\!\frac{b_0}{2}\,p
    \delta(z-z_\alpha)\!\left[\! 1\! - \!\tanh\left(\!5
    \frac{\left|\vec{r}-\vec{r}_\alpha\right|-b_0}{b_0}\!\right)\!\right] ,
  \end{split}
\end{equation}
where $\vec{r}_\alpha$ and $z_\alpha$ respectively denote the in-plane and 
$z$ coordinates of pinning center $\alpha$ , and $p \geq 0$ represents the 
pinning potential strength. 
Columnar defects are set up by first randomly selecting the in-plane
coordinates $\vec{r}_\alpha$, and then placing identical potential wells 
\eqref{eq:pinningwells} at all $z$ positions $z_\alpha$.  
Henceforth, all lengths are measured relative to the pinning potential width 
$b_0$, and energies are measured in units of $\epsilon_0 b_0$ with 
$\epsilon_0 = (\phi_0/4\pi\lambda_{ab})^2$, where $\phi_0=hc/2e$ is the 
magnetic flux quantum.

\subsection{Langevin Molecular Dynamics}
\label{sec:lang-molec-dynam}

We analyze the vortex kinetics by means of a Langevin mole\-cular dynamics 
algorithm. 
As mentioned above, we discretize space in the $z$ direction (along the 
external magnetic field) into layers, whose spacing corresponds to the unit
cell size $c_0$ along the crystal's $c$ direction~\cite{Das2003,Bullard2008}. 
The effective forces acting on each flux line element are then computed as
spatial derivatives of the discretized version of the Hamiltonian 
(\ref{eq:hamiltonian}).
Correspondingly, we discretize and numerically solve the overdamped 
Langevin equation 
\begin{equation}
  \label{eq:langevin}
  \eta\,\frac{\partial\vec{r}_i(t,z)}{\partial t} =
  -\frac{\delta H[\vec{r}_i(t,z)]}{\delta\vec{r}_i(t,z)}+\vec{f}_i(t,z) \, ,
\end{equation}
with the Bardeen-Stephen viscous drag parameter $\eta$ by direct temporal
integration~\cite{Brass1989}. 
Note that all fast degrees of freedom ({\em e.g.}, phonons etc.) are captured 
by thermal stochastic forces, which are modeled as uncorrelated Gaussian white 
noise with vanishing mean $\left<\vec{f}_i(t,z)\right> = 0$.
We impose the Einstein relation 
\begin{equation}
  \label{eq:noise}
  \left<\vec{f}_i(t,z)\vec{f}_j(s,z')\right> = 
  2\eta\,k_{\rm B}T\,\delta_{ij}\delta(t-s)\delta(z-z')
\end{equation}
to guarantee that the system asymptotically reaches thermal equilibrium with
a canonical probability distribution $\propto e^{- H / k_{\rm B} T}$.  
The flux lines are moving in a three-dimensional space with periodic boundary 
conditions in the $xy$ plane and free boundary conditions in the $z$ direction. 
We set the system size to ($16/\sqrt{3}) \lambda_{ab} \times 8\lambda_{ab}$,
since this aspect ratio ensures that the system equilibrates to form a 
hexagonal Abrikosov lattice of the present flux lines in the absence of 
disorder and driving current.

\subsection{Material Parameters}
\label{sec:material-parameters}

We select our material parameters guided by the actual values for the ceramic 
high-$T_c$ type-II superconducting compound YBa$_2$Cu$_3$O$_7$ (YBCO). 
All simulation distances are measured in units of the pinning center radius, 
which we set to $b_0=35\AA$. 
We choose the spacing between layers in the $c$ direction equal to this basic
microscopic scale, $c_0=b_0$. 
The in-plane London penetration depth is $\lambda_{ab}=34b_0\approx 1200\AA$, 
and the superconducting coherence length $\xi_{ab}=0.3b_0\approx 10.5\AA$. 
This emulates the high anisotropy of YBCO, with an effective mass anisotropy 
ratio $\Gamma^{-1}=1/5$. 
All simulation energies are measured in units of $\epsilon_0 b_0$, where 
$\epsilon_0$ is the line energy per unit length: 
$\epsilon_0\approx1.92\cdot 10^{-6}\operatorname{erg}/\operatorname{cm}$. 
The vortex line tension energy scale becomes
$\tilde{\epsilon}_1/\epsilon_0\approx 0.189$. 
The depth of the pinning potential wells is fixed at $p/\epsilon_0=0.05$. 
We perform our numerical runs at temperatures $T=10\ldots 40\,$K, with 
$T=10\,$K corresponding to $k_{\rm B} T / \epsilon_0 b_0=0.002$ in our 
simulation units.  
Finally, we set the Bardeen-Stephen viscous drag coefficient 
$\eta=\phi_0^2/2\pi\rho_nc^2\xi_{ab}^2 \approx10^{-10} \operatorname{erg}
\cdot\operatorname{s} / \operatorname{cm}^2$ to one (for the normal-state 
resistivity of YBCO near $T_c$, $\rho_n\approx 500\,\mu\Omega\,$cm, see 
Table~1 in Ref.~\cite{Abdelhadi1994}); thus the simulation time steps are 
fixed by the fundamental temporal unit 
$t_0=\eta b_0/\epsilon_0\approx 18\,$ps.
In the following, all times are measured in simulation time steps, i.e., in 
units of $t_0$.

\subsection{Simulation Protocol}
\label{sec:relax-simul-prot}

Our aim is to impose experimentally realizable initial conditions to bridge 
theoretical simulations and experimental work. 
To this end, we investigate sudden changes in the magnetic field or 
temperature. 
A magnetic field quench is implemented in our simulations by adding flux 
lines at random positions in the system (up-quench) or removing randomly
selected vortices from the system (down-quench). 
We start with straight flux lines positioned at random in the sample, and let 
the system relax up to an initial relaxation time $r=10^5$.  
We then instantaneously increase the magnetic field by adding 5 straight flux
lines to the 16 originally present and essentially relaxed vortices; or 
instantaneously decrease the field by removing 5 random lines from 21 relaxed 
initial vortices. 
Subsequently, we let the resulting system relax up to a waiting time $s$, at 
which a snapshot of the system is taken; the waiting time $s$ is measured 
after the quench. 
At later times $t>s$, different two-time correlation functions are evaluated.
Since $r$ is much larger than the different waiting times $s$ and measurement
time $t$, it is crucial to take this elapsed time period into account when 
considering the initial lines that have relaxed for a duration $r$.
We therefore introduce the times $\sigma=r+s$ and $\Gamma=r+t$. 

When the external magnetic field is altered in experimental samples, flux lines
enter or leave the sample from its boundaries. Since our numerical study is 
limited to a fairly small number of vortices, introducing physical surfaces 
through the removal of our implemented periodic boundary conditions would 
render our system dominated by the ensuing boundary effects. We rather consider 
our simulation domain as a representative sample of the bulk region in a much 
larger physical system. Also, the formation or deletion of vortices from the 
sample's boundaries occurs on time scales much faster than considered in this 
study of comparatively slow relaxation kinetics.

Temperature quenches are accomplished by letting the initial configuration of 
$N=16$ flux lines relax up to a similar initial relaxation time ($t=10^5$)
at which the temperature is suddenly raised or lowered. 
Throughout our analysis of magnetic field and temperature quenches, the number
of layers is $L=640$ and the number of pinning sites per layer is fixed at 
$N_D/L=1710$.

\subsection{Measured Quantities}
\label{sec:measured-quantities}

We measure various two-time correlation functions, some of which provide 
insight into the local features of the fluctuating vortex lines, such as their
{\em height-height autocorrelations}
\begin{equation}
\label{eq:heightheightcorrelation}
  C(t,s)=\left< \left[\vec{r}_{i,z}(t)-\overline{\vec{r}}_i(t)\right]
  \cdot \left[\vec{r}_{i,z}(s)-\overline{\vec{r}}_i(s)\right]\right> \, ,
\end{equation}
where $\vec{r}_{i,z}(t)$ are the in-plane coordinates of line $i$ at layer $z$
at time $t$, and $\overline{\vec{r}}_i(t)= \left<\vec{r}_{i,z}(t)\right>$ is 
the mean lateral position of line $i$ at time $t$. 
Here, observable quantitities are averaged over all line elements as well as 
many noise histories and disorder realizations: The brackets 
$\langle a \rangle$ indicate the average over noise and disorder 
realizations, where $a=\frac{1}{NL} \sum_{i=1}^N \sum_{z=1}^L a_{i,z}$ denotes 
the average of the local variable $a_{i,z}$ over all line elements present in 
the system. 
$C(t,s)$ thus measures the thermal fluctuations of line elements around their 
corresponding mean lateral line position. 
We remark that the term `height-height autocorrelation function' for the 
transverse vortex fluctuations derives from viewing the flux lines as 
fluctuating one-dimensional interfaces described by a unique local height 
function that measures the deviation of $\vec{r}_{i,z}(t)$ from the respective 
line's mean.

The global structure of the flux line configuration is analyzed using the 
two-time {\em mean-square displacement}
\begin{equation}
\label{eq:meansqdisplacement}
  B(t,s)=\left<\left[\vec{r}_{i,z}(t)-\vec{r}_{i,z}(s)\right]^2\right> \, .
\end{equation}
This correlation function measures the average square distance between the 
position of a line element at time $s$ and the position of that same line 
element at a later time $t$. 
Both two-time correlation functions were also analyzed in 
Refs.~\cite{Pleimling2011} and \cite{Dobramysl2013}, where Monte Carlo and
Langevin molecular dynamics simulations, respectively, were performed to 
characterize the complex non-equilibrium relaxation of initially straight and 
randomly placed vortex lines in the presence of point-like disorder.  
Thermal spatial fluctuations can also be quantified by means of the vortex 
line radius of gyration 
\begin{equation}
\label{eq:gyration_radius}
  r_g(t)=\sqrt{\left<\left[\vec{r}_{i,z}(t)
  -\overline{\vec{r}}_i(t)\right]^2\right>} \ , 
\end{equation}
i.e., the root mean-square displacement from the lines' mean lateral position 
at time $t$. 

Another one-time observable that we measure in our simulations is the 
fraction $\varphi(t)$ of pinned line elements when disorder is present. 
To this end, we first find the number of line elements that reside within a
fixed small cut-off distance from a certain pinning center at a given time 
$t$.
Then we divide this count by the total number of line elements present in the 
sample to yield $\varphi(t)$. 
Throughout our study, the cut-off distance is set equal to the pinning center 
radius, $r_c=b_0$. 
The total number of line elements can by computed by multiplying the number of 
layers $L=640$ in the $z$ direction with the number $N$ of vortex lines 
present in the system.

\subsection{Finite-Size Effects}
\label{sec:finite size}

\begin{figure}[b]
\centering
\subfloat{\includegraphics[width=0.98\columnwidth]{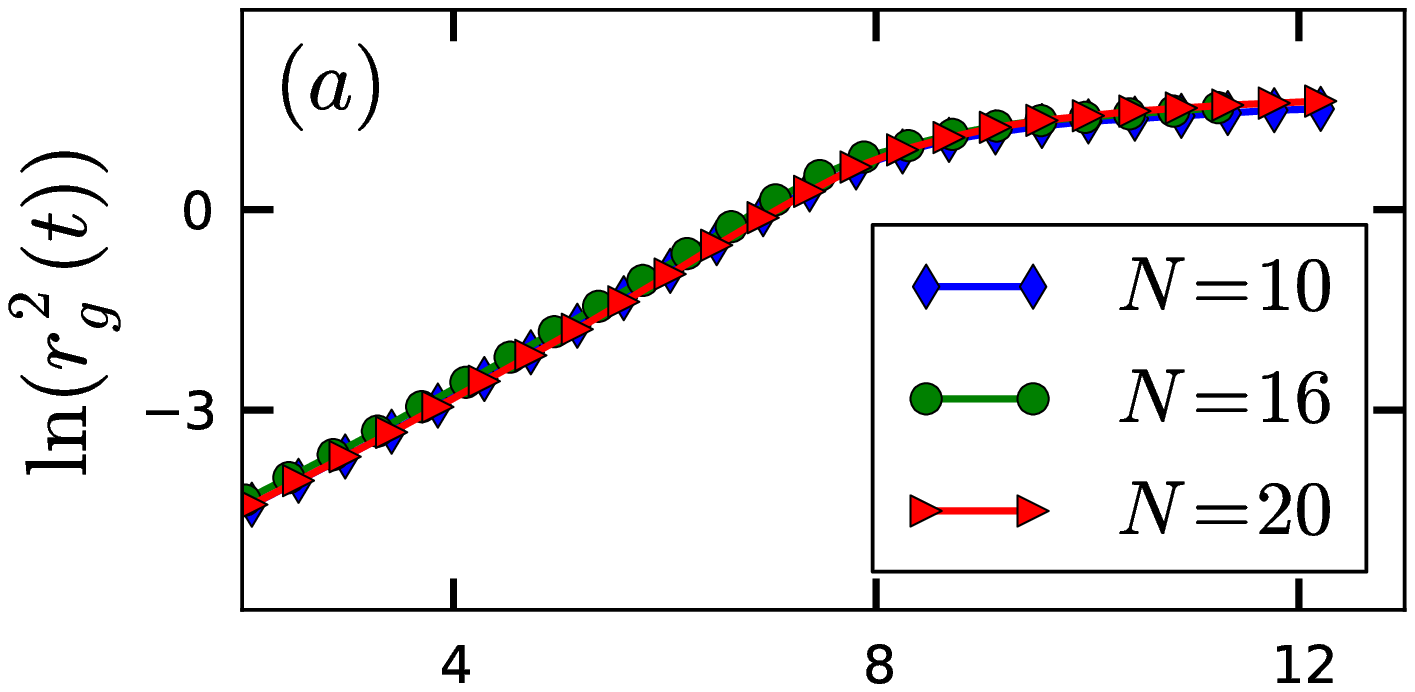}
  \label{fig:finite-0ad-int-point-rg}} \\ [-1.3ex] 
  \subfloat{\includegraphics[width=0.98\columnwidth]{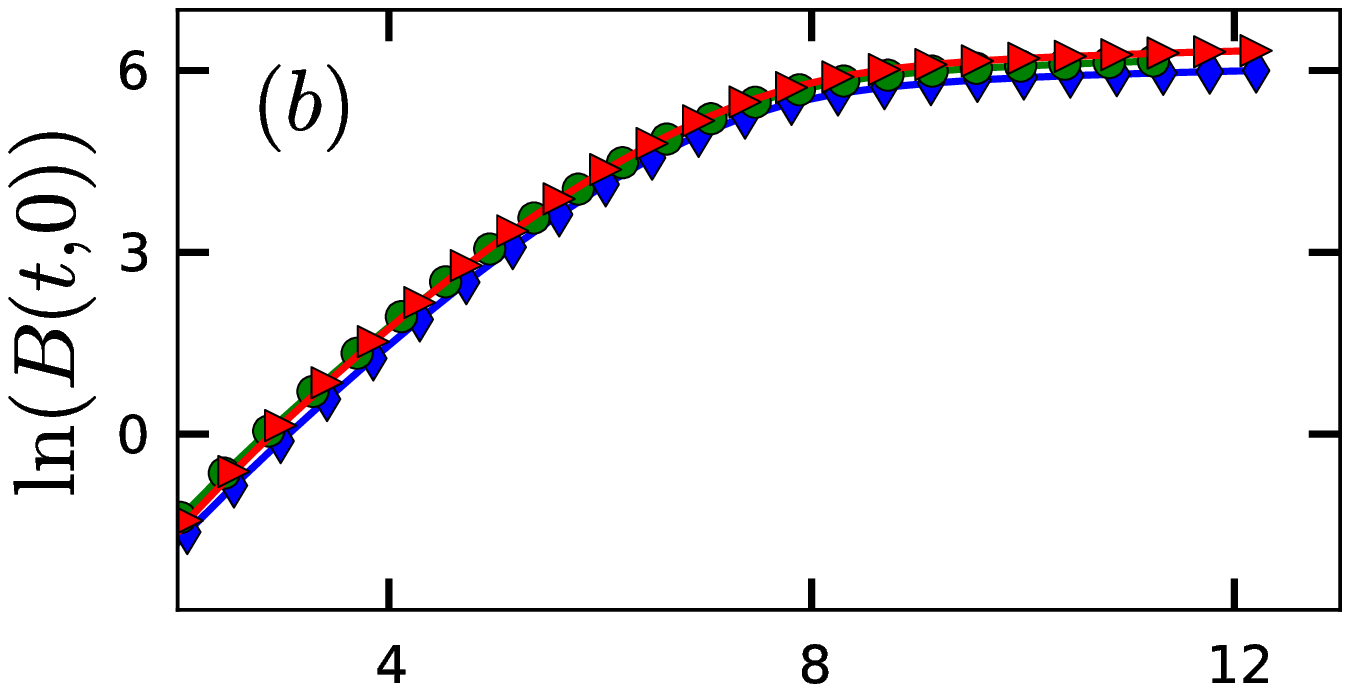}
  \label{fig:finite-0ad-int-point-b}} \\ [-1.4ex]
  \subfloat{\includegraphics[width=0.98\columnwidth]{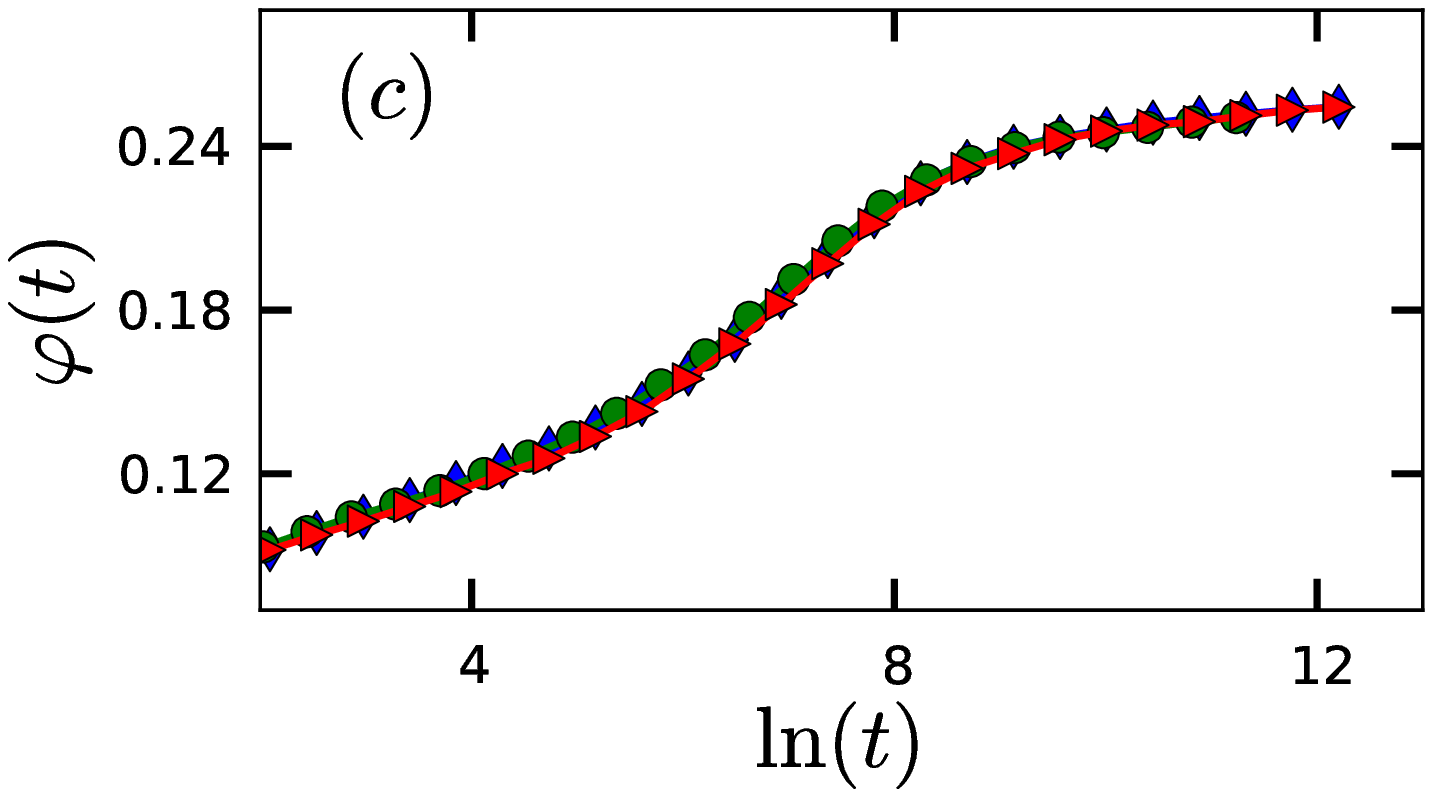}
  \label{fig:finite-0ad-int-point-pp}}
  \caption{(Color online) Relaxation of the (a) gyration radius, 
    (b) mean-square displacement, and (c) fraction of pinned line elements for 
    (dilute) systems with different number of interacting vortex lines in the 
    presence of point-like pinning centers, averaged over 100 disorder and 
    noise realizations.}
\label{fig:finite-0ad-int-point}
\end{figure}
In our work, we restrict ourselves to a low magnetic field or vortex density
regime, and only consider systems composed of between $N=10$ and $N=21$ 
vortices.  
This naturally constitutes a tiny system compared to typical experimental 
samples of disordered type-II superconductors in external magnetic fields. 
It is therefore crucial to analyze the effects of this small vortex density, 
especially when we remove lines as the magnetic field sudden decreases.  
We consequently ran simulations in systems with different numbers of flux 
lines, but keeping the density of pinning centers the same in all cases.

As demonstrated in Fig.~\ref{fig:finite-0ad-int-point}, the data for the 
vortex radius of gyration $r_g(t)$, mean-square displacement $B(t,0)$, and 
the fraction $\varphi(t)$ of pinned line elements that correspond to systems 
of different number of flux lines collapse onto a common curve. 
Therefore, we conclude that these differently-sized systems follow similar 
relaxation processes confirming that finite-size effects do not play a 
prominent role in this computational study of dilute systems of flux lines in 
a dissipative medium. 
It is worth noting that the physics in denser vortex systems could of course
be quite different.

\section{Temperature Quenches}
\label{sec:temp}

We first consider sudden changes in the ambient temperature of the system of 
interacting vortex lines, with the goal to analyze the effects of 
instantaneous increases in temperature on the system's relaxation dynamics. 
At $t=10^5$ in our simulation, we instantaneously increase the temperature 
from $T=0.002$ to $T=0.008$ in a system of $N=16$ interacting flux lines in 
the presence of either point or columnar pinning centers. 
We then compare the relaxation of the radius of gyration $r_{g}(t)$ and the 
fraction of pinned line elements $\varphi(t)$ when the temperature stays fixed
with the scenario wherein it is suddenly raised. 
We remark that the error bars displayed in some of the figures in this section,
however small they appear, always represent the standard error of the mean of 
the computed quantities. 
We observe that the relaxation of these observables in the considered systems 
can invariably be mapped to an exponential function of the form 
$a\,e^{-t/\tau}+c$, where $\tau$ determines a characteristic relaxation time. 
The measured values for the relaxation times $\tau$ in different scenarios are 
summarized in Table~\ref{table:rel-time}. 
We note that the exponential relaxation of observables in our vortex model 
system following sudden temperatures quenches implies that physical aging 
features would not be discernible on time scales larger than the relaxation 
time $\tau$.
\begin{table}[!htb]
\centering
\resizebox{\linewidth}{!}{%
\begin{tabular}{|c|c|c|c|c|}
\cline{1-5} \multirow{2}{*}{} Defect & $T=0.002$ & $T=0.008$ 
& \multicolumn{2}{c|}{$T=0.002 \to 0.008$} \\ \cline{2-5} 
type & \multicolumn{3}{c|}{$r_g$} & $\varphi$ \\ \hline
\multicolumn{1}{|l|}{(a)\,Point-like} & $5.9\cdot 10^4$ & $5.3\cdot 10^4$  
& $3.5\cdot 10^4$ & $3.1\cdot 10^4$ \\ \hline
\multicolumn{1}{|l|}{(b)\,Columnar} & $3.5\cdot 10^4$ & $4.9\cdot 10^4$ &
\begin{tabular}[c]{@{}l@{}}$\tau_1=1.6\cdot 10^3$ \\ $\tau_2=5.7\cdot 10^4$ 
\end{tabular} & $3.3\cdot 10^4$ \\ \hline
\end{tabular}}
\caption{Measured characteristic relaxation times for the gyration radius $r_g$
  and the fraction of pinned line elements $\varphi$ in systems of interacting 
  vortex lines with (a) point-like and (b) columnar defects (data averaged over
  1000 realizations).}
\label{table:rel-time}
\end{table}

\begin{figure}
  \centering
  \subfloat{\stackinset{r}{0.3cm}{b}{0.9cm}
  {\includegraphics[height=3.6cm]{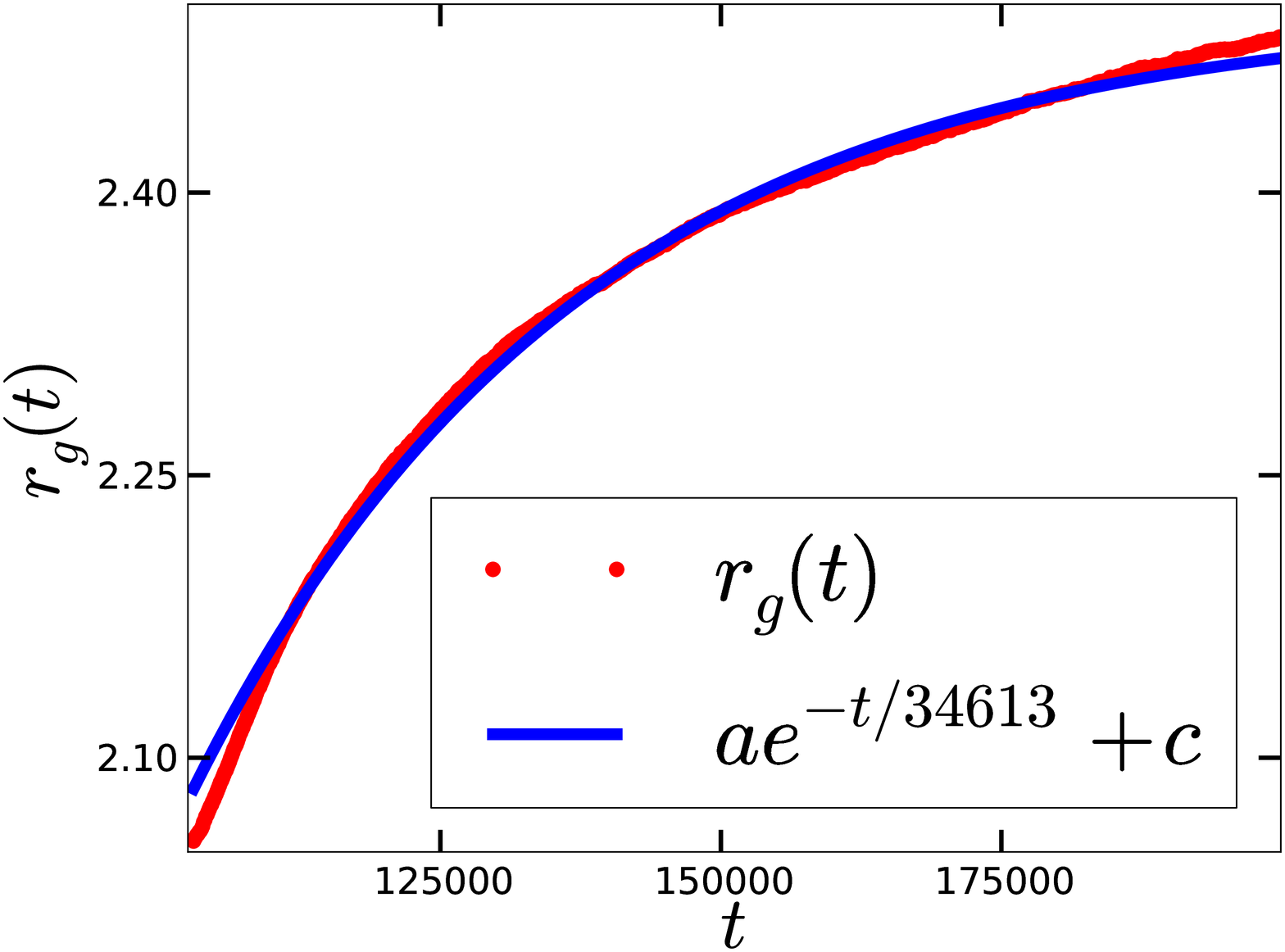}}
  {\includegraphics[width=0.98\columnwidth]{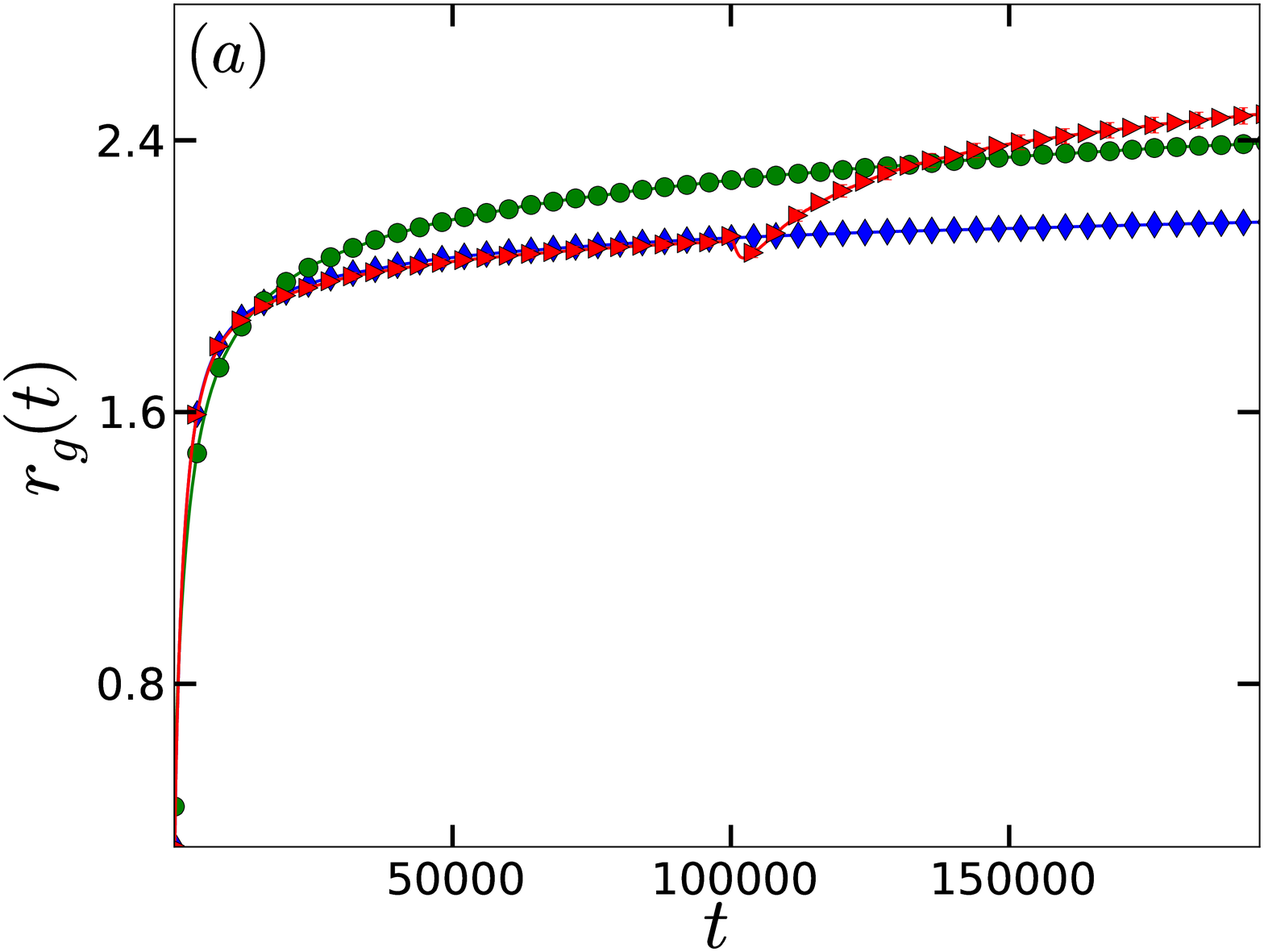}
  \label{fig:temp-up-int-point-rg}}} \\ 
  \subfloat{\includegraphics[width=0.98\columnwidth]{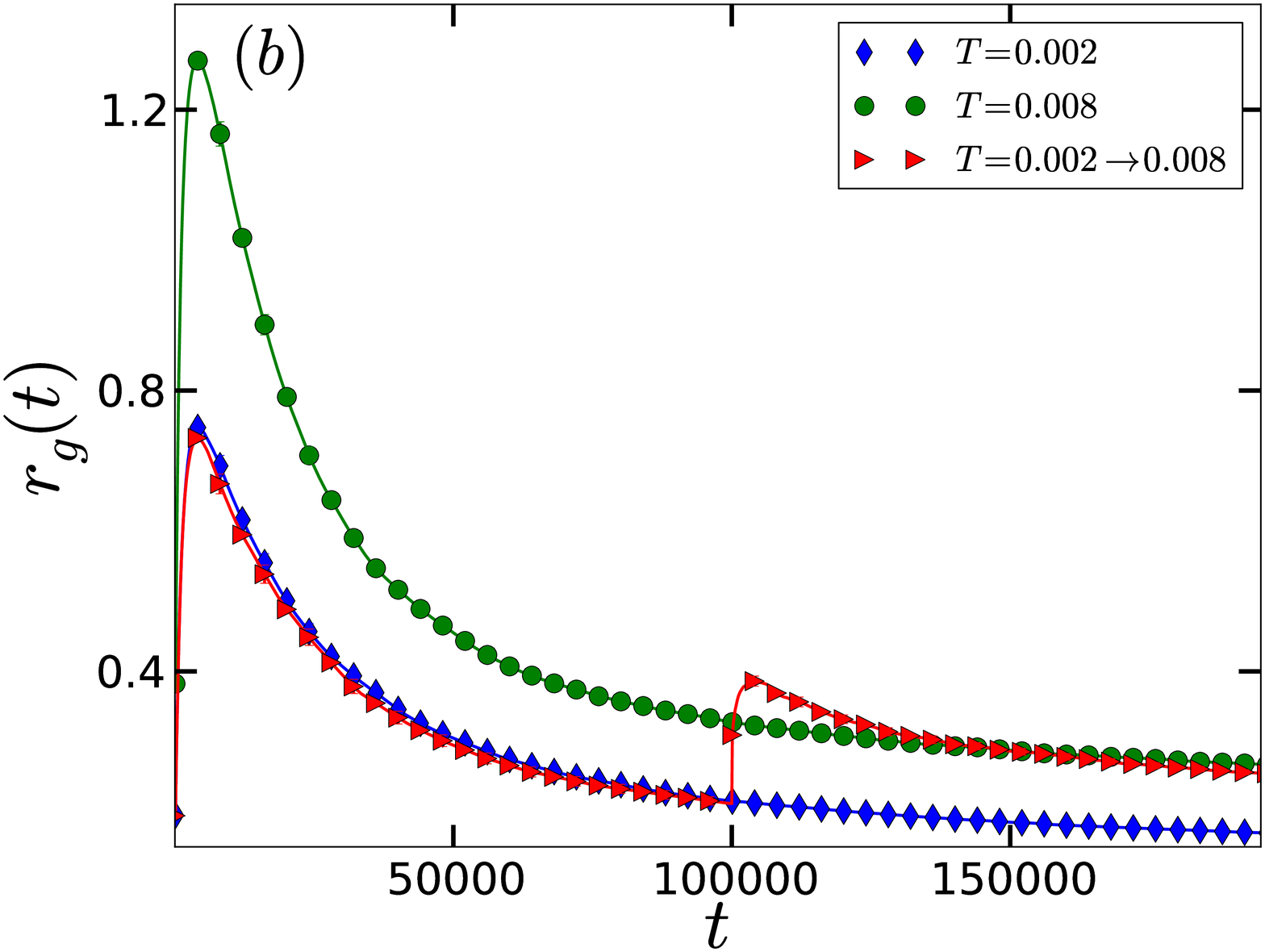}
  \label{fig:temp-up-int-col-rg}}
  \caption{(Color online) Relaxation of the gyration radius in systems of 
    interacting vortex lines with (a) point disorder or (b) columnar defects, 
    when the temperature is either held fixed or instantaneously raised (data 
    averaged over 1000 realizations). 
    The inset in (a) shows the gyration radius following the sudden temperature
    increase in the presence of point disorder, and a fit to an exponential 
    function.}	
\label{fig:temp-up-int-point-col-rg}
\end{figure}

In systems with point-like disorder, the measured relaxation time of the 
radius of gyration at high temperature is shorter than that at low temperature,
see Table~\ref{table:rel-time}. 
This confirms that thermal fluctuations facilitate the roughening of vortices 
in the presence of point-like disorder. 
Following a sudden increase in temperature, the radius of gyration first
decreases a little as a few flux line segments become thermally depinned, and
subsequently increases monotonically, as shown in 
Fig.~\ref{fig:temp-up-int-point-rg}. 
We checked that this increase in $r_g(t)$ is well described by an exponential 
function of the form $a\,e^{-t/\tau}+c$, with relaxation time 
$\tau=3.5\cdot 10^4$ (the other parameters are $a\approx-1.73$ and 
$c\approx-0.09$) as observed in the inset in 
Fig.~\ref{fig:temp-up-int-point-rg}. 
Thus, as expected, point defects enhance thermal wandering of flux lines in 
this disordered landscape.

In contrast, systems with columnar defects actually require a longer 
relaxation time for the radius of gyration at higher temperatures than that at 
lower temperatures, see Table~\ref{table:rel-time}, asserting that thermal 
fluctuations resist the straightening of vortex lines as they become localized
at columnar pinning centers.
When we instantaneously increase the temperature in this system, the gyration 
radius after the quench shows a marked fast increase, even beyond the 
relaxation curve for a system at constant higher temperature, followed by a 
much slower decrease, see Fig.~\ref{fig:temp-up-int-col-rg}.
The initial increase in $r_g$ can likely be attributed to the creation of 
double-kinks in the system that are not similarly generated in relaxation 
processes at fixed temperature, while the final decrease is due to the decay of 
these metastable configurations. 
A double-kink refers to a flux line configuration where the vortex is 
simultaneously pinned to two adjacent columnar defects. 
This is a long-lived state which will eventually decay to either a free vortex,
or, more probably, to a straight flux line bound to a single defect column.
We find that both the initial increase and the final decrease in the gyration 
radius again fit an exponential of the form $a\,e^{-t/\tau}+c$, but with 
different relaxation times (and signs of the coefficient $a$): 
In Fig.~\ref{fig:temp-up-int-col-rg}, the fast relaxation time from the initial
non-equilibrium steady state at $T=0.002$ to the metastable state with vortex 
double-kinks is $\tau_1=1.6\cdot 10^3$, while the much longer relaxation time 
from this intermediate state to the final non-equilibrium steady state at 
$T=0.008$ is $\tau_2=5.7\cdot 10^4$. 
Note that the data for the gyration radius for both the temperature up-quench 
scenario as well as at a fixed high temperature in 
Figs.~\ref{fig:temp-up-int-point-rg} and \ref{fig:temp-up-int-col-rg} will 
eventually approach the same curve in the long-time limit, which has not yet 
been reached in our simulation time window.

\begin{figure}
  \centering
  \subfloat{\includegraphics[width=0.98\columnwidth]{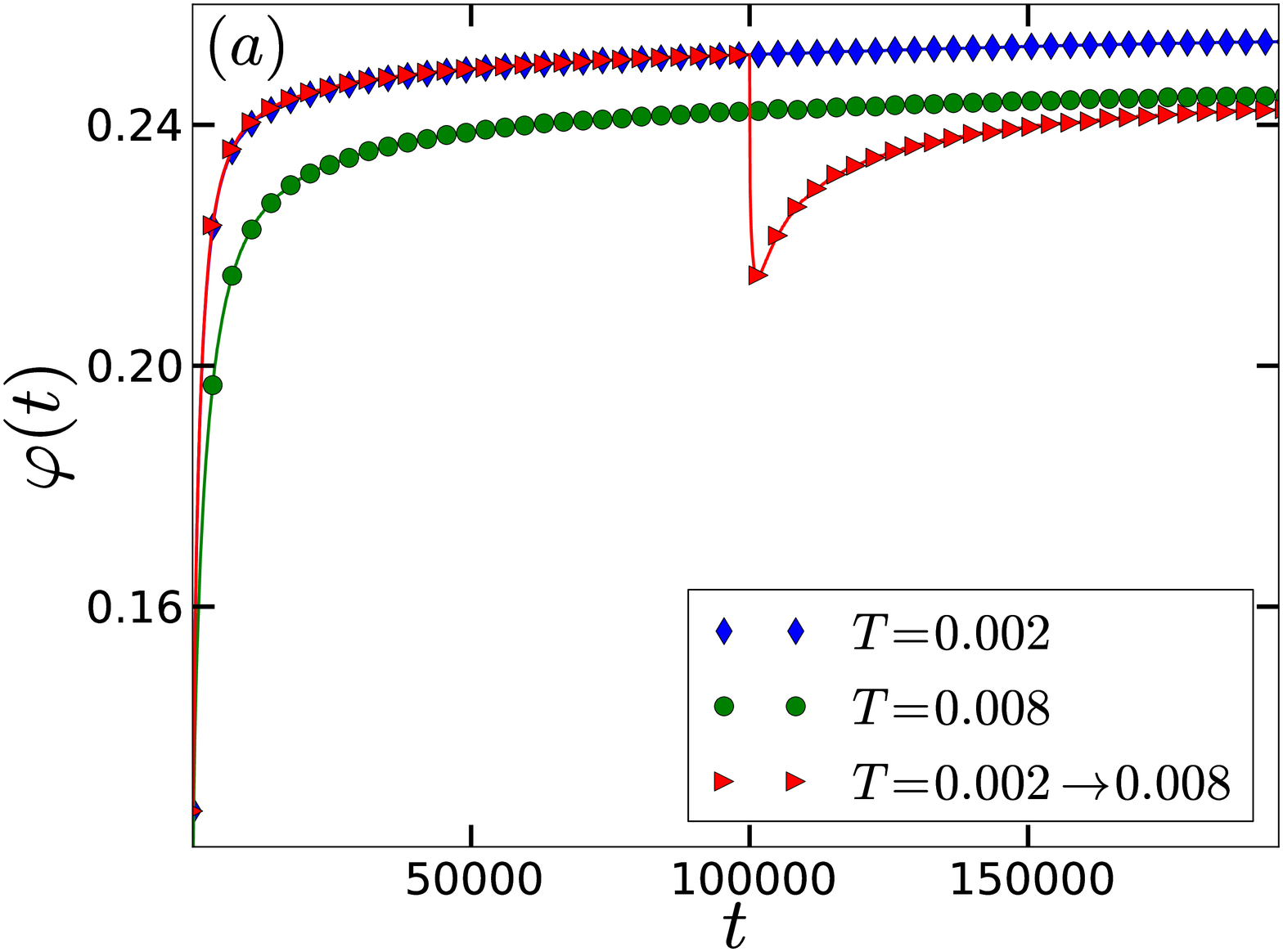}
  \label{fig:temp-up-int-point-pp}} \\ 
  \subfloat{\includegraphics[width=0.98\columnwidth]{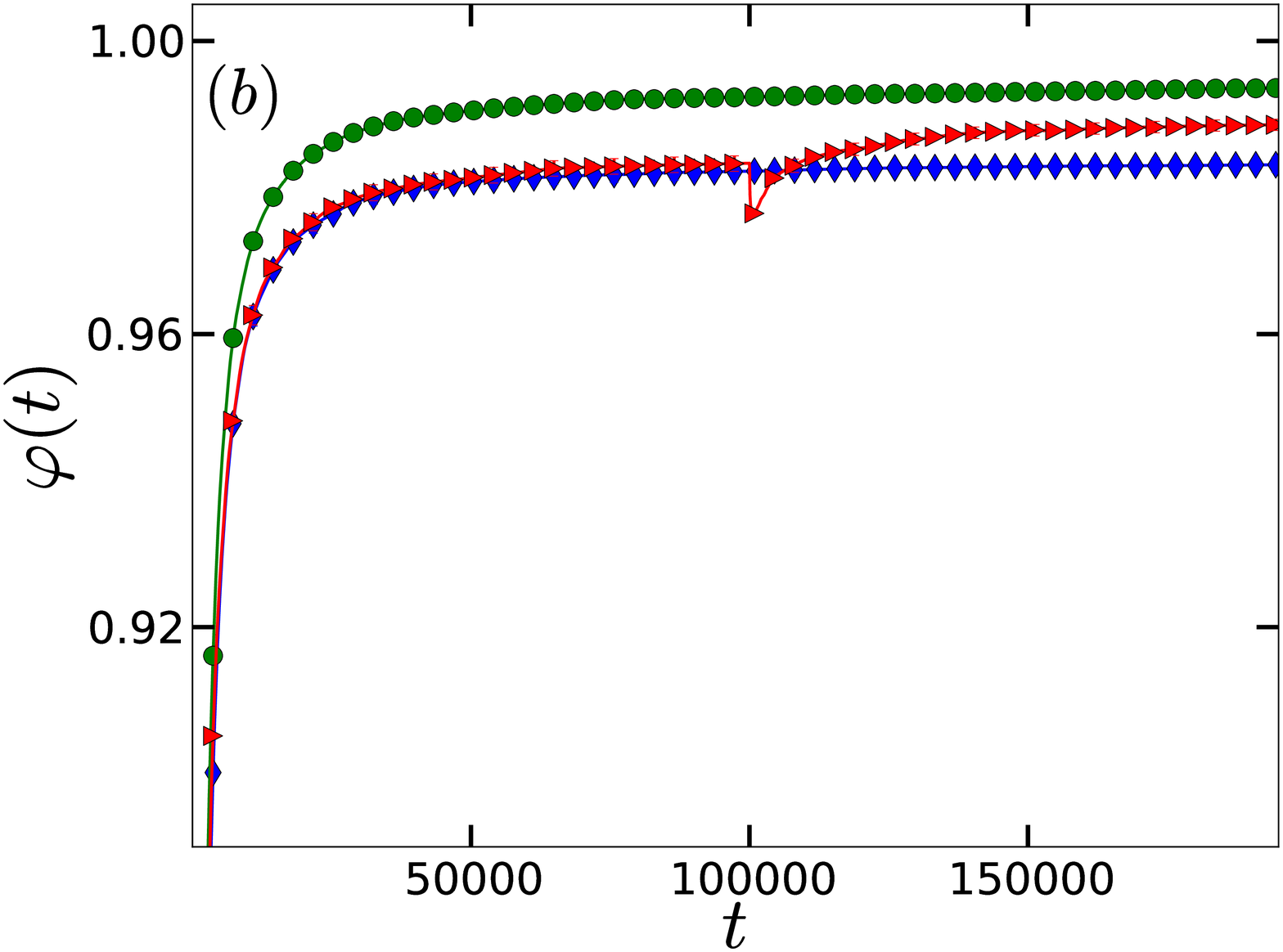}
  \label{fig:temp-up-int-col-pp}}
  \caption{(Color online) Relaxation of the fraction of pinned line elements in
    systems of interacting vortex lines with (a) point disorder or (b) columnar
    defects, when the temperature is either held fixed or instantaneously 
    raised (data averaged over 1000 realizations).}	
\label{fig:temp-up-int-point-col-pp}
\end{figure}
The differences between the systems' responses to temperature up-quenches in 
the presence of point and columnar pinning centers, respectively, can be 
attributed to the very distinct (de-)pinning behavior that we shall again
discuss later in the context of sudden magnetic field changes.  
When the temperature is suddenly raised in systems with uncorrelated point 
disorder, some line elements become thermally depinned.
The vortices are initially straightened owing to their line tension.
Subsequently an increasing number of line elements finds favorable pinning
sites, causing a monotonic growth in their transverse spatial fluctuations
measured by $r_g$. 
In samples with columnar defects, following a temperature up-quench some line 
elements migrate away from certain columnar pinning centers, which renders them
subject to thermal wandering.
Yet they soon become pinned to neighboring defect lines causing the subsequent 
decrease in the lateral spatial vortex line fluctuations, akin to systems 
relaxing at fixed temperature.

This scenario is confirmed by the relaxation data for the fraction of pinned 
line elements $\varphi(t)$ in Fig.~\ref{fig:temp-up-int-point-col-pp}. 
As observed in Fig.~\ref{fig:temp-up-int-point-pp}, in the presence of 
point disorder only a minority ($\approx 25\%$) of line elements are actually
pinned right before the temperature quench. 
A sudden increase in temperature releases a significant fraction of these 
pinned line elements causing a decrease in $\varphi$ as well as the small dip
in $r_g$.
Later, the vortices become attached to and hence stretched between point-like
pins causing a subsequent increase in $\varphi$, accompanied with a monotonic 
increase in the lines' spatial fluctuations. 

\begin{figure}
  \centering
  \includegraphics[width=0.98\columnwidth]{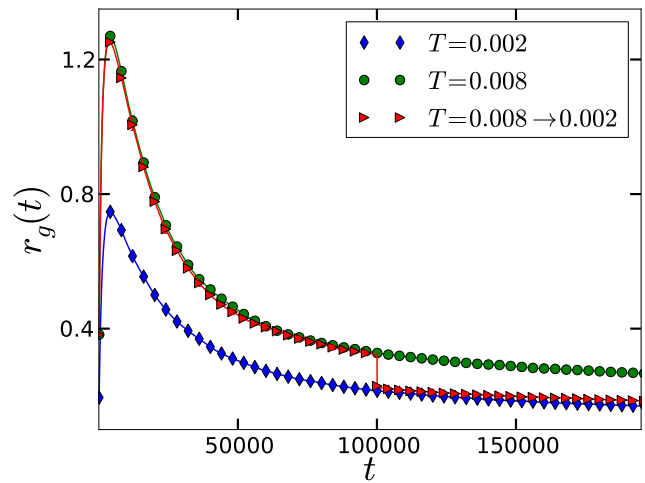}
  \caption{(Color online) Relaxation of the gyration radius in a system of 
    interacting vortex lines with columnar pinning centers, when the 
    temperature is held fixed or instantaneously lowered (data averaged over 
    1000 realizations).}
\label{fig:temp-down-int-col-rg}
\end{figure}
In the presence of columnar defects, on the other hand, 
Fig.~\ref{fig:temp-up-int-col-pp} shows that almost all line elements are bound
to pinning centers just before the temperature quench is applied.
The sudden temperature increase causes a tiny number of these line elements to 
depin from their localizing columns.
However, they are soon to be pinned again, but partly in metastable double-kink
configurations.
This explains the non-monotonic pattern we observe in the spatial fluctuations 
with a rapid increase of $r_g$ followed by a slow decrease. 
Finally, the fraction of pinned line elements $\varphi(t)$ following the 
temperature quench in the presence of both types of disorder once again fits an 
exponential of the form $a\,e^{-t/\tau}+c$, yet with different relaxation 
times: $\tau=3.1\cdot 10^4$ for point disorder, whereas $\tau=3.3\cdot 10^4$ 
for columnar defects, see Table~\ref{table:rel-time}.
This confirms that enhanced thermal fluctuations at elevated temperatures 
resist the straightening of flux lines as they attempt to bind to columnar 
defects; in stark contrast, thermal effects facilitate line roughening when 
vortices become pinned to uncorrelated point-like disorder.

We also analyzed the effects of a sudden drop in temperature, where we 
reduced the system's ambient temperature at $t=10^5$ from $T=0.008$ to 
$T=0.002$ in our simulation runs.
The subsequent vortex relaxation kinetics displays some noticeable differences
as compared with the previous up-quenches: 
Following the temperature decrease, the gyration radius decreases and the 
fraction of pinned line elements increases in samples with both types of 
disorder, and non-monotonic behavior is observed in neither of these 
quantities, as depicted for $r_g$ in Fig.~\ref{fig:temp-down-int-col-rg}. 
When the temperature suddenly drops, more line elements simply become pinned 
to either point-like or extended defects, which causes a concomitant decrease 
in the transverse spatial fluctuations of the vortex lines.

\section{Magnetic Field Quenches}
\label{sec:mag}

\subsection{Non-interacting Vortex Lines without Disorder}
\label{sec:noint-nodis}

\begin{figure}
 \centering
 \subfloat
 {\includegraphics[width=0.98\columnwidth]{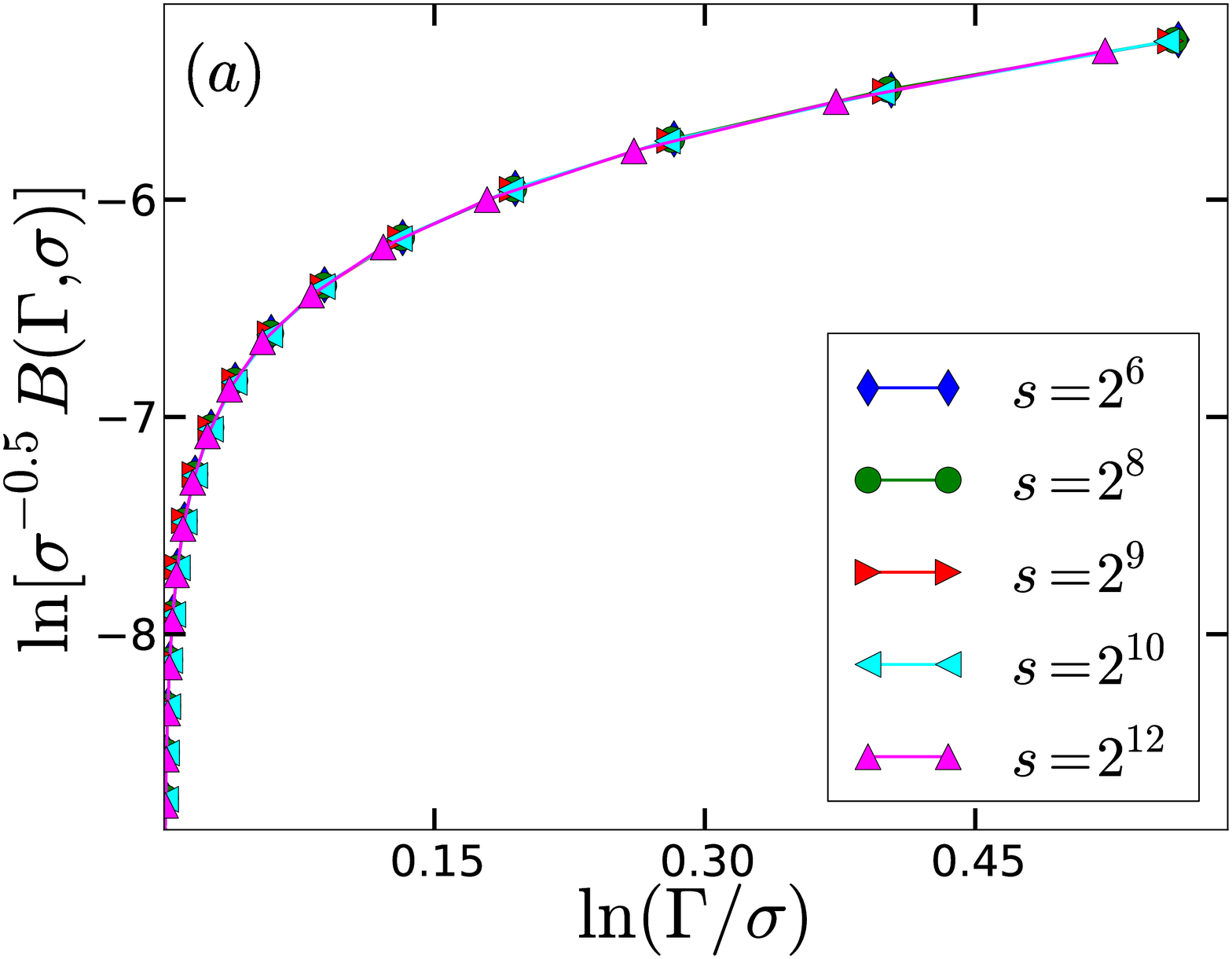}
 \label{fig:noquench-noint-nodis-b}}\\
 \subfloat
 {\includegraphics[width=0.98\columnwidth]{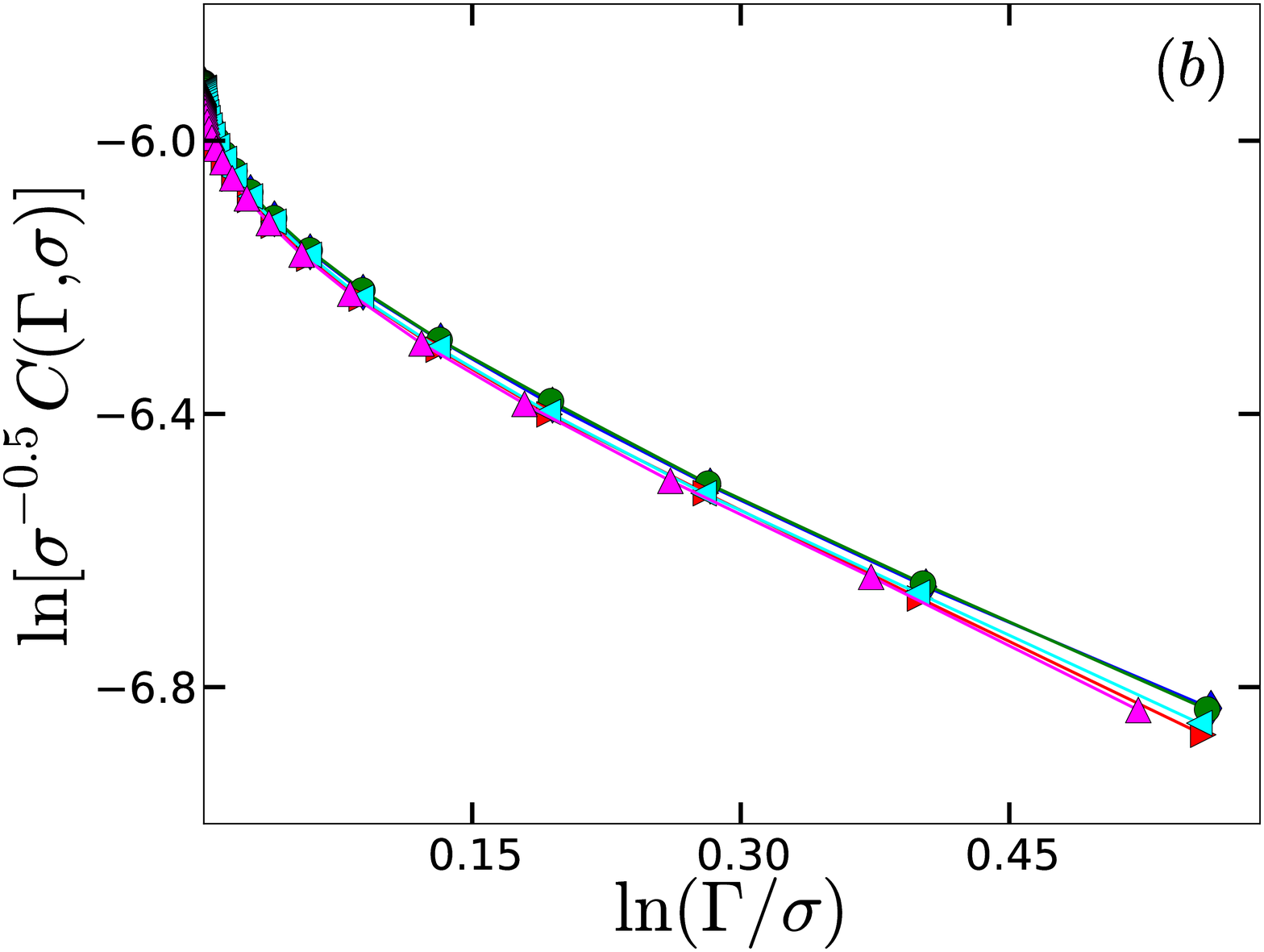}
 \label{fig:noquench-noint-nodis-c}}
 \caption{(Color online) Relaxation of (a) the mean-square displacement and (b)
   the height-height autocorrelation function for non-interacting vortices in 
   the absence of disorder and for fixed vortex density (data averaged over 800
   realizations).}	
\label{fig:noquench-noint-nodis}
\end{figure}
We now proceed to study the effects of sudden changes in the external magnetic
field, {\em i.e.}, vortex density, while keeping the ambient temperature fixed 
at $T=0.002$ in all simulation runs for the various systems described in this 
section. 
The dynamics of non-interacting directed lines in the absence of disorder can
be directly mapped to the one-dimensional Edwards-Wilkinson interface growth 
model (which is in turn equivalent to a free noisy diffusion equation). 
The two-time height-height autocorrelation function 
\eqref{eq:heightheightcorrelation} in the correlated regime is found to 
be~\cite{Rothlein2006} 
\begin{equation}
  \label{eq:1}
  C(t,s)=C_0s^{1/2}\left(\left[\frac{t}{s}+1\right]^{1/2}-
  \left[\frac{t}{s}-1\right]^{1/2}\right)\, .
\end{equation}
Indeed, if a simple aging scenario applies, one expects the general scaling 
form for the height-height autocorrelation function \cite{Henkel2010}
\begin{equation}
  \label{eq:agingscaling}
  C(t,s)=s^b f_C(t/s)
\end{equation}
with a scaling function $f_C$.
In the free diffusive Edwards-Wilkinson regime, the scaling exponent is 
$b=1/2$~\cite{Rothlein2006,Pleimling2011}. 
In our Langevin molecular dynamics study, we first analyze this case of free
non-interacting vortex lines and start with the fixed magnetic field scenario
to validate our numerical code. 
Since the present flux lines relax for a long initial time $r=10^5$, which is 
much larger than the waiting times $s$ and measurement time $t$, we need to 
take this elapsed time $r$ into account when analyzing the behavior of these 
lines, and therefore new times $\sigma=r+s$ and $\Gamma=r+t$ are introduced.
The two-time mean-square displacement \eqref{eq:meansqdisplacement} follows a 
similar scaling form to \eqref{eq:agingscaling}. 
As shown in Fig.~\ref{fig:noquench-noint-nodis}, our data for both 
$B(\Gamma,\sigma)$ and $C(\Gamma,\sigma)$ indeed satisfy dynamical scaling for 
all explored waiting times with the predicted scaling exponent $b=1/2$, after 
properly accounting for the long initial relaxation time $r$, see 
Sec.~\ref{sec:relax-simul-prot}. 

When we follow our magnetic field quench scenario, and suddenly reduce the 
vortex density after letting the system relax for an extended time period, we 
obtain the very same relaxation results for both $B(\Gamma,\sigma)$ and 
$C(\Gamma,\sigma)$, which is to be expected since the lines are not 
interacting. 

Similarly, when we instantaneously increase the magnetic field, 
$B(\Gamma,\sigma)$ and $C(\Gamma,\sigma)$ measured for those vortices that were 
present initially are not significantly different from those quantities 
obtained at fixed flux density, as the relaxation processes for these 
non-interacting vortex lines are statistically the same in both situations. 
However, the subpopulation of newly added lines displays a similar, but less 
perfect collapse for $C(t,s)$ than that displayed in 
Fig.~\ref{fig:noquench-noint-nodis-c} since these lines were not present for 
the previous long relaxation time $r$, and thus are not yet as relaxed as the 
initial lines. 
On the other hand, $B(t,s)$ obtained just for the added lines exhibits perfect
dynamical scaling with the scaling exponent $b=1/2$.

In summary, we confirm the expected Edwards-Wilkin\-son aging scaling for the 
non-equilibrium relaxation dynamics of non-interacting flux lines in the 
absence of disorder, regardless of sudden changes in their density. 
The following subsections highlight the effects of disorder on the relaxation 
dynamics of non-interacting, thus statistically independent vortex lines.

\subsection{Non-interacting Flux Lines with Point Disorder}
\label{sec:noint-point}

Keeping the repulsive vortex-vortex interactions switched off, we now introduce 
point defects into the system to study the effects of this type of spatially
uncorrelated disorder on the relaxation dynamics.

\begin{figure}[b]
 \centering
 \includegraphics[width=0.98\columnwidth]{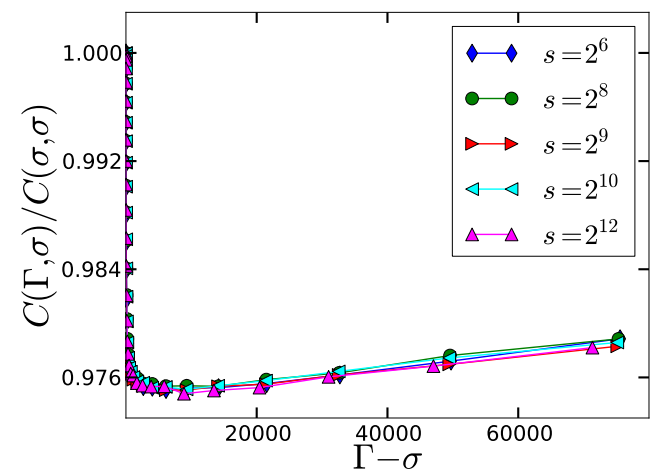}
 \caption{(Color online) Relaxation of the normalized height-height 
   autocorrelation function in a system of non-interacting flux lines subject 
   to point-like disorder, when the number of flux lines stays fixed (data 
   averaged over 800 realizations).}	
\label{fig:noquench-noint-point-c}
\end{figure}
First, we consider the situation of fixed magnetic field (vortex density), and
study the same dynamical quantities as in the previous section, see also
Refs.~\cite{Pleimling2011,Dobramysl2013}.
The mean-square displacement $B(\Gamma,\sigma)$ does not display dynamical 
scaling with $b=1/2$ as in the disorder-free case.
Instead, dynamical scaling ensues for $B(\Gamma,\sigma)$ with the aging
exponent $b \approx 0.725$, {\em c.f.} Fig.~6(d) in Ref.~\cite{Dobramysl2013}. 
In contrast, dynamical scaling is not observed for the height-height 
autocorrelation function $C(\Gamma,\sigma)$. 
As shown in Fig.~\ref{fig:noquench-noint-point-c}, the data for all waiting 
times $s$ display identical time evolution, since the system has in fact fully 
equilibrated owing to the long relaxation period $r=10^5$, and time translation 
invariance is restored.
The height autocorrelation function shows a similar increase for all waiting 
times indicating that flux line elements keep exploring their surroundings in
search of favorable pinning sites, which causes the lines to continuously 
roughen due to the uncorrelated nature of the point-like disorder in the 
system. 

\begin{figure}
 \centering
 \subfloat{
 \includegraphics[width=0.97\columnwidth]{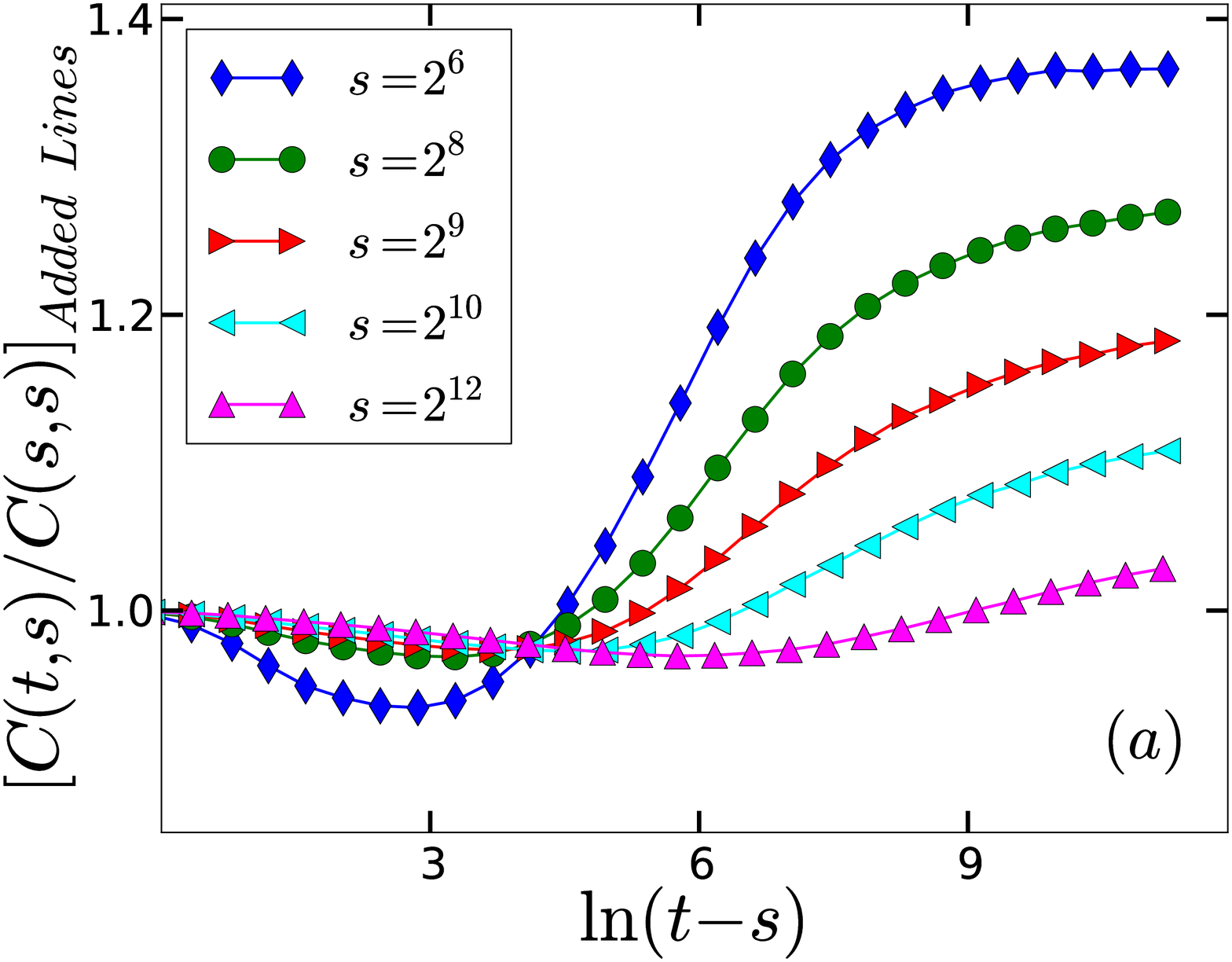} 
 \label{fig:upquench-noint-point-cad}} \\
 \subfloat{
 \includegraphics[width=0.97\columnwidth]{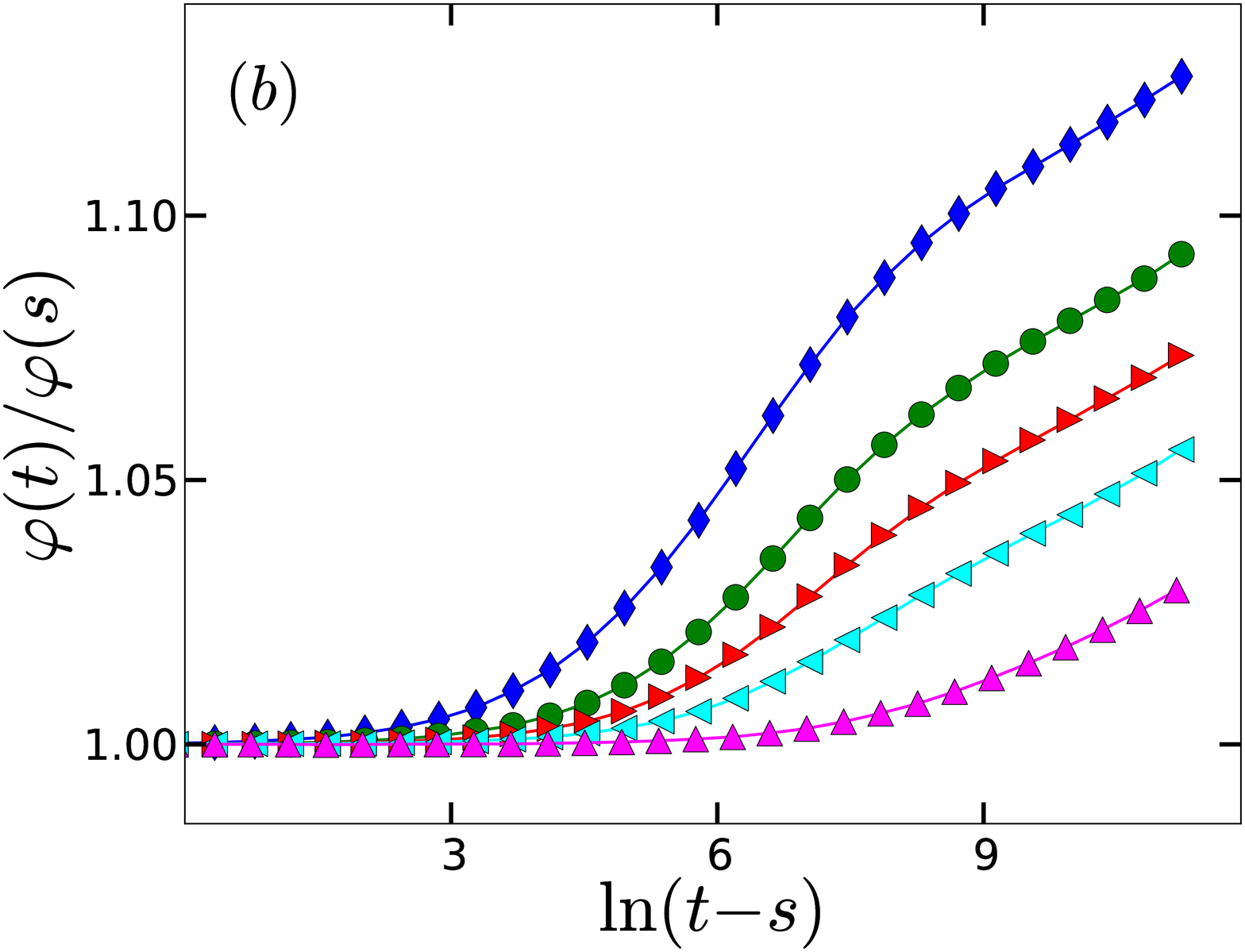}
 \label{fig:upquench-noint-point-pp}}
 \caption{(Color online) Relaxation of (a) the normalized height-height 
   autocorrelation function of the newly added flux lines, and (b) the fraction
   of pinned line elements in a system of non-interacting vortices subject to 
   point disorder, following a sudden increase in the magnetic field or vortex 
   density (data averaged over 800 realizations).}
\label{fig:upquench-noint-point}
\end{figure}
The results for a sudden magnetic field decrease (vortex density down-quench) 
turn out to not significantly differ from the situation with constant field. 
This is readily attributed to the fact that the lines are not interacting, and 
thus the remaining vortices after the quench are not at all affected by the 
removal of some lines from the sample.

More distinct features are observed upon an instantaneous increase of the 
magnetic field:
First of all, the mean-square displacement $B(\Gamma,\sigma)$ of the initial 
lines does not display dynamical scaling anymore, but the data (not shown) 
reveal that this vortex subpopulation has rather recovered time translation 
invariance, since they all collapse onto a single curve when plotted against 
$\Gamma-\sigma=t-s$, while the added line subpopulation displays neither of
these straightforward features. 
The height-height autocorrelation function $C(\Gamma,\sigma)$ of the initial 
lines shows time translation invariance, akin to the data in 
Fig.~\ref{fig:noquench-noint-point-c}. 
On the other hand, $C(t,s)$ for the added lines, depicted in 
Fig.~\ref{fig:upquench-noint-point-cad}, displays an initial decrease followed 
by an increase, with the minimum shifted to later $t-s$ values for longer 
waiting times $s$. 
The decrease happens when the added lines start noticing the presence of the
pinning centers and become pinned to nearby defects, which decreases their 
transverse fluctuations.
The subsequent increase is caused by the flux lines exploring more of the 
sample and thus becoming rougher as they optimize their pinning configurations. 

Hence, the behavior of the added lines is highly influenced by disorder as 
demonstrated by their height autocorrelations. 
A one-time observable that also highlights the effects of the sudden increase 
in the magnetic field is the fraction of pinned vortex line elements $\varphi$. 
In the presence of point disorder, this fraction as shown in 
Fig.~\ref{fig:upquench-noint-point-pp} displays a two-step increase, which can 
be attributed to the pinning dynamics of the newly added magnetic flux lines, 
as discussed in the context of Fig.~\ref{fig:upquench-noint-point-cad}.

\subsection{Non-interacting Vortices with Columnar Pins}
\label{sec:noint-col}

We proceed to change the type of pinning centers present in the system to 
columnar defects, which are linearly extended pinning sites that are simulated 
as potential wells aligned along the $z$ direction, and compare their effects 
on the relaxation kinetics of non-interacting vortices to that of uncorrelated
point disorder. 

\begin{figure*}
 \centering
 \subfloat{\includegraphics[width=0.98\columnwidth]{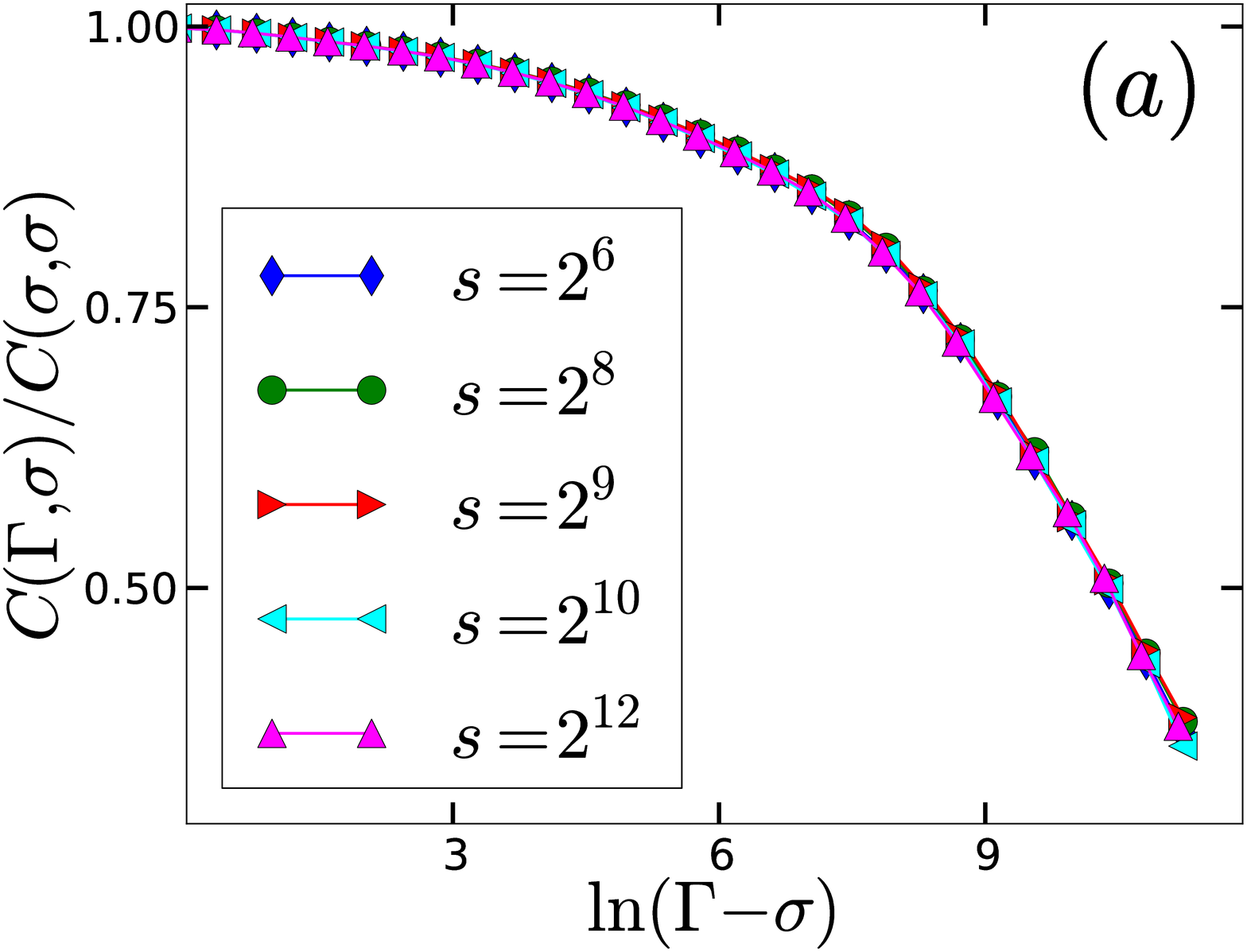}
 \label{fig:noint-col-noquench-c}} \
 \subfloat{\includegraphics[width=0.98\columnwidth]{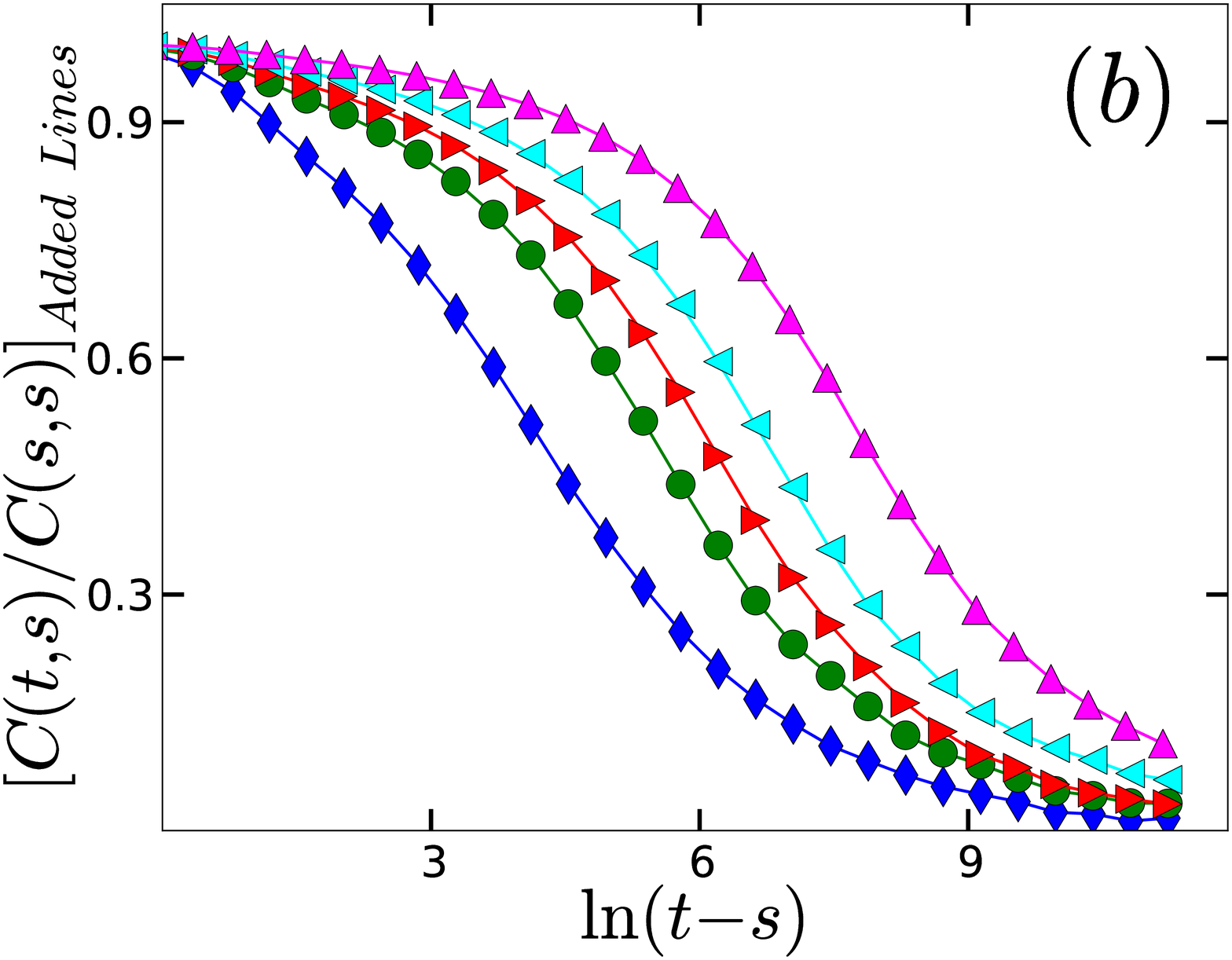}
 \label{fig:noint-col-upquench-cad}} \\
 \subfloat{\includegraphics[width=0.98\columnwidth]{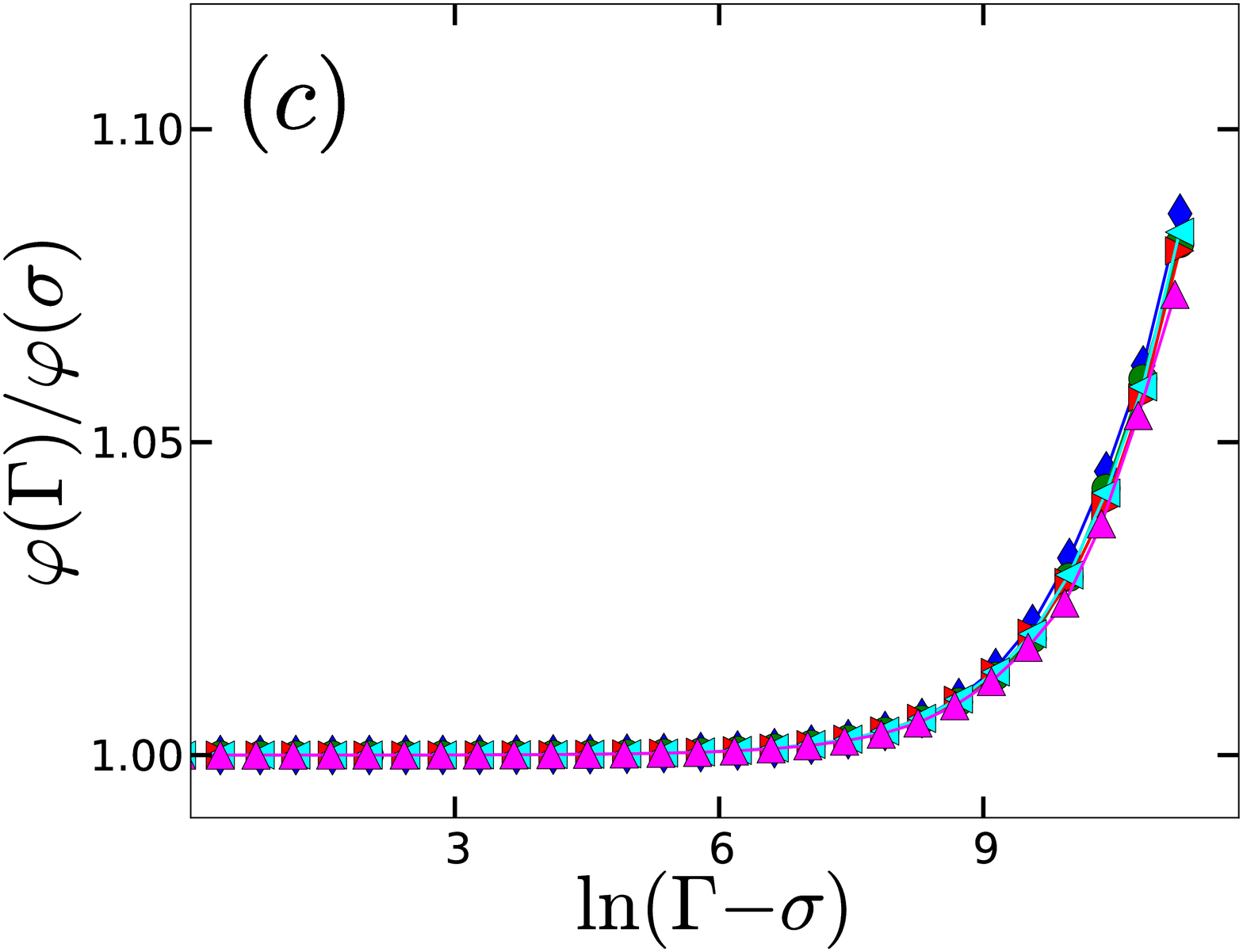}
 \label{fig:noint-col-noquench-pp}} \
 \subfloat{\includegraphics[width=0.98\columnwidth]{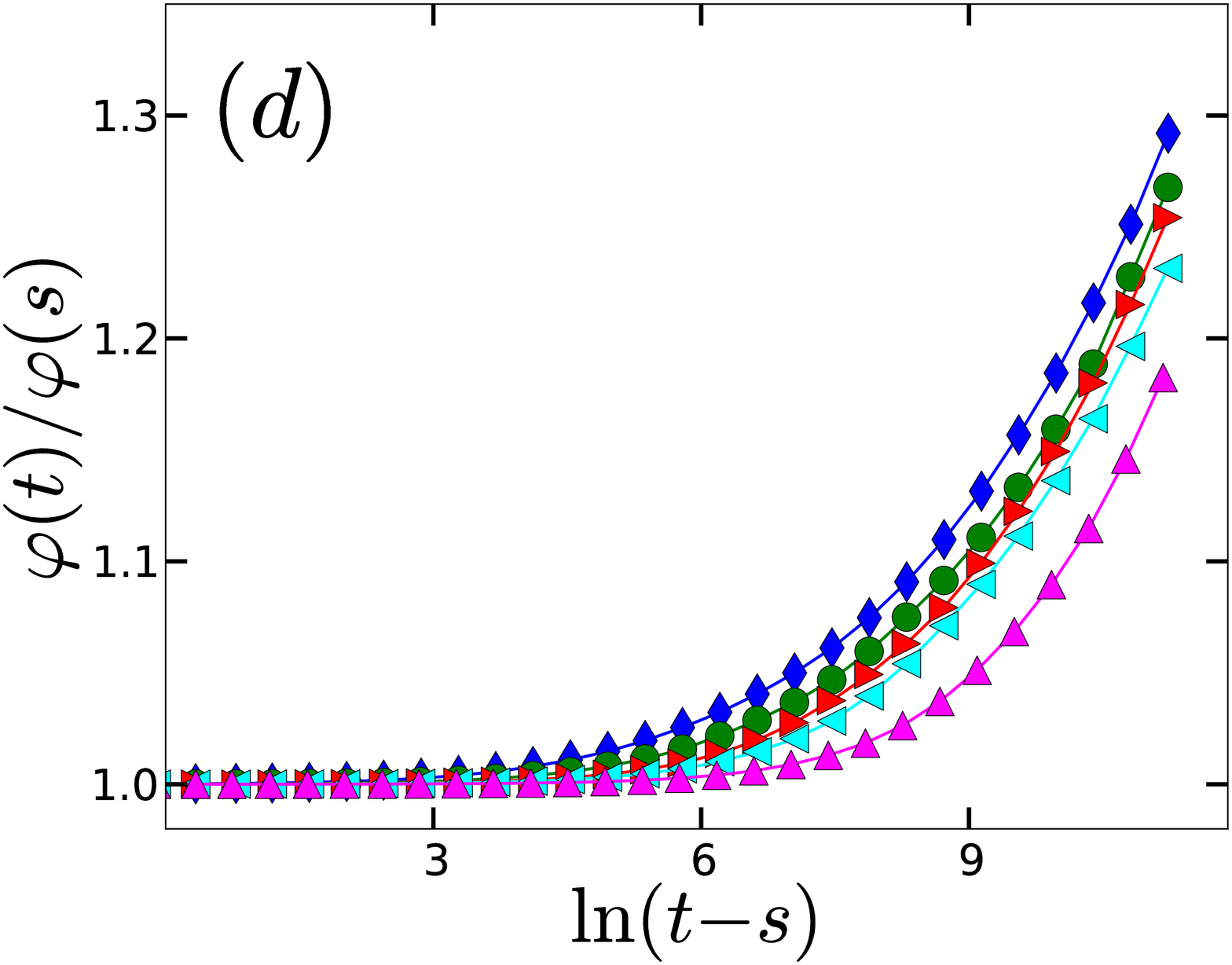}
 \label{fig:noint-col-upquench-pp}}
 \caption{(Color online) Relaxation of various observables in a system of 
   non-interacting flux lines with columnar defects, (a,c) when the vortex 
   density remains fixed; (b,d) after a sudden increase in the magnetic field:
   (a,b) normalized height-height autocorrelation function, (c,d) fraction of 
   pinned line elements (data averaged over 800 realizations).} 
\label{fig:noint-col-noquench-upquench}
\end{figure*}
We again first hold the magnetic field constant and consider the different 
two-time correlation functions studied above. 
The mean-square displacement $B(\Gamma,\sigma)$ is found to display dynamical 
scaling with the aging exponent $b=1/2$, similar to 
Fig.~\ref{fig:noquench-noint-nodis-b} in the absence of disorder. 
It is worth recalling that when flux lines are completely pinned to columnar 
defects, their mean-square displacement becomes constant, of the order of 
$b_0$. 
We therefore conclude that the dynamics of $B(\Gamma,\sigma)$ in the accessible
time window is clearly dominated by the still unattached lines in the system.
The height-height autocorrelation function $C(\Gamma,\sigma)$ does not display 
dynamical scaling with any fixed scaling exponent. 
However, when considering the normalized $C(\Gamma,\sigma)$ with respect to the
initial value $C(\sigma,\sigma)$ as depicted in 
Fig.~\ref{fig:noint-col-noquench-c}, we observe that the data for all different 
waiting times $s$ perfectly collapse onto a single curve, which indicates that 
time translation invariance is restored and the system is well equilibrated.
Next, we measure the fraction of pinned line elements in the system with 
columnar defects, see Fig.~\ref{fig:noint-col-noquench-pp}. 
The data for different waiting times collapse onto a single graph, where the 
fraction of pinned line elements stays constant for an extended time period and 
only later, at $\ln(\Gamma-\sigma) \approx 7$, starts increasing.
This indicates that $\Gamma \approx 1096+\sigma$ represents a characteristic 
time scale when pinning to columnar defects becomes prominent in the system.

When vortices are removed from this system of non-in\-teracting flux lines with 
columnar defects, the measured mean-square displacement, height 
autocorrelations, and fraction of pinned line elements all yield identical 
results to the situation when there is no sudden magnetic field change, as is 
to be expected in the absence of correlations. 

Upon introducing additional flux lines into the system, the mean-square 
displacement of both the initial and the added lines subpopulations shows 
dynamical scaling for all waiting times with the same exponent $b=1/2$ as in 
the fixed density and field down-quench scenarios.
This confirms that sudden changes in the vortex density do not alter the global 
dynamics of non-interacting flux lines in the presence of columnar defects. 
The autocorrelation function $C(\Gamma,\sigma)$ for the initially present lines 
turns out very similar to that at fixed density, see 
Fig.~\ref{fig:noint-col-noquench-c}, whereas the data corresponding to the 
newly added lines cannot be collapsed according to a simple aging scaling form 
when plotted against the time ratio $t/s$. 
However, the graphs $C(t,s)$ for the added lines for various $s$ do not just 
fall onto a single curve when plotted against $t-s$, as demonstrated in 
Fig.~\ref{fig:noint-col-upquench-cad}.
This indicates that this subpopulation of lines is not yet equilibrated for 
the range of waiting times $s$ employed here.  
As the additional lines are inserted into the sample, depending on the elapsed 
waiting time $s$, they require different time periods $t$ to explore the system
and become attached to columnar pins.  
This is corroborated in the time evolution of the fraction of pinned line 
elements $\varphi$ for different waiting times in 
Fig.~\ref{fig:noint-col-upquench-pp}, which also does not yield data collapse.

The flux lines present in the sample, especially vortices that were suddenly
added to the system, start exploring the surrounding pinning centers sooner for 
shorter waiting times, whence one observes an earlier increase in the fraction 
of pinned line elements for smaller $s$. 
For the long waiting times in Fig.~\ref{fig:noint-col-upquench-pp} such as 
$s=2^{12}$, the fraction of pinned line elements is fixed at a plateau value 
for an extended duration because the vortices already had enough time to relax
towards favorable configurations after the field up-quench. 
The increase in $\varphi(t)$ happens at a much later time indicating that the 
line elements might be migrating between neighboring columnar pinning centers 
to become pinned to whole defect lines, whereupon their lateral line 
fluctuations diminish. 
The overall trend in the presence of columnar defects is that the dynamics is 
dominated by the few lines that remain unattached to any pins, and the 
tendency of each flux line to become fully localized to a single columnar 
pinning center, which in turn straightens the vortex lines. 

The preceding discussions highlight the influence of different disorder on 
non-interacting vortex matter and the relaxation dynamics following magnetic 
field quenches. 
We expect that quite distinct relaxation kinetics will be observed when mutual
vortex interactions are added to our consideration and collective behavior 
ensues, as will be explored in the following subsections.

\subsection{Interacting Vortex Lines without Disorder}
\label{sec:int-nodis}

\begin{figure}
 \centering
 \subfloat{
 \includegraphics[width=0.97\columnwidth]{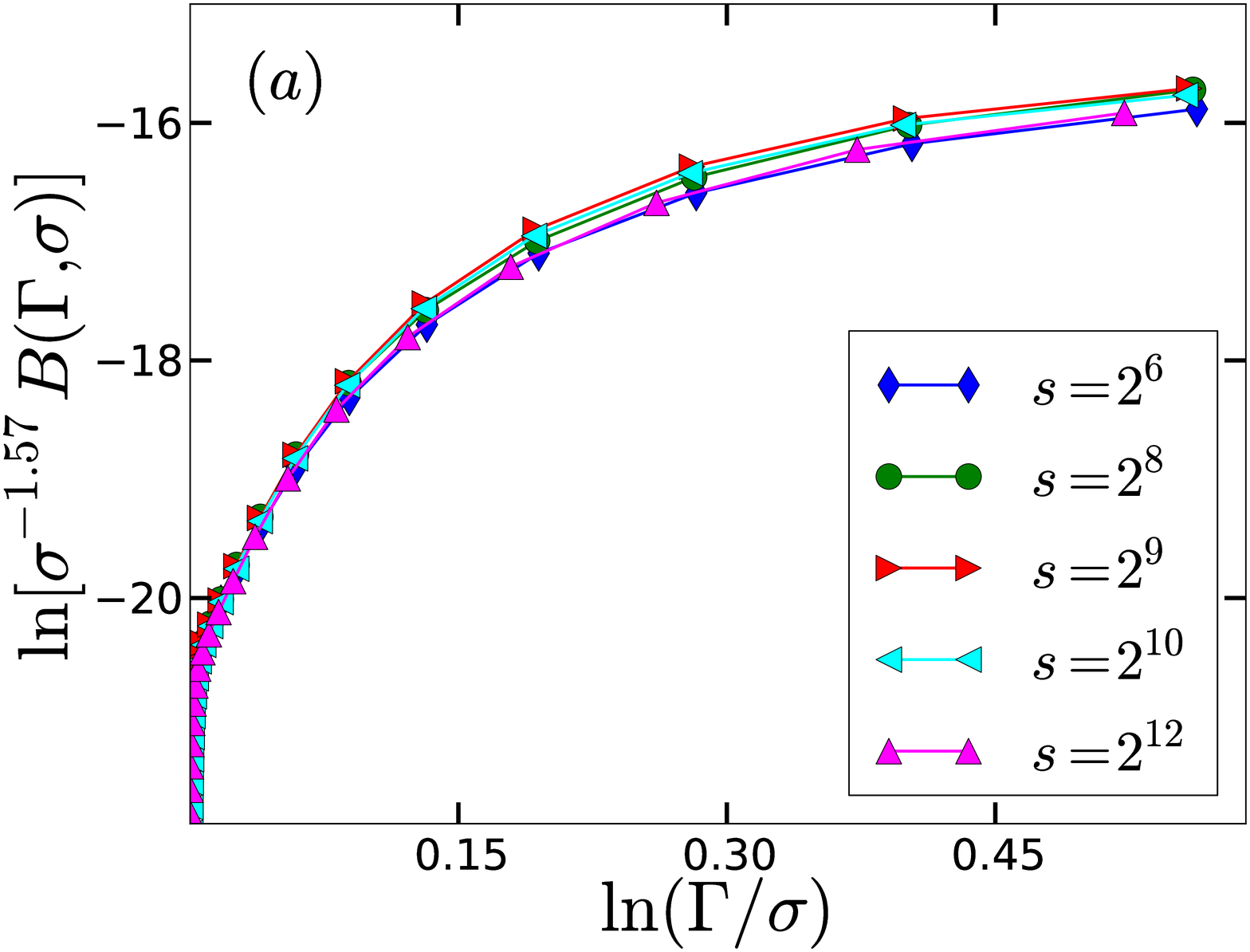}
 \label{fig:int-nodis-noquench-b}} \\
 \subfloat{
 \includegraphics[width=0.97\columnwidth]{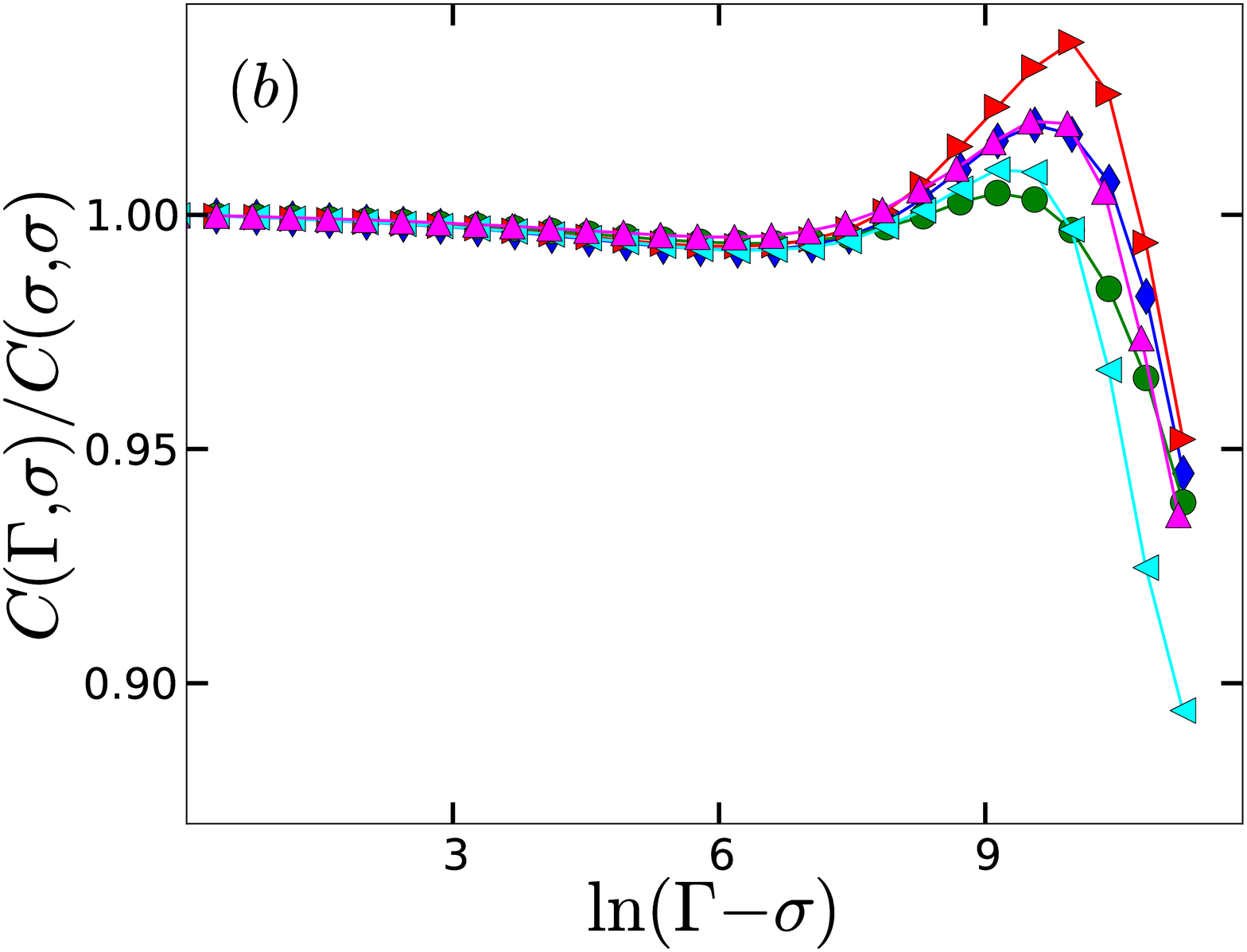}
 \label{fig:int-nodis-noquench-c}}
 \caption{(Color online) Relaxation of (a) the mean-square displacement and 
   (b) the normalized height-height autocorrelation function in a system of 
   interacting flux lines in the absence of disorder, when the number of flux 
   lines stays fixed (data averaged over 800 realizations).}	
\label{fig:int-nodis-noquench}
\end{figure}

\begin{figure}
 \centering
 \subfloat{
 \includegraphics[width=0.97\columnwidth]{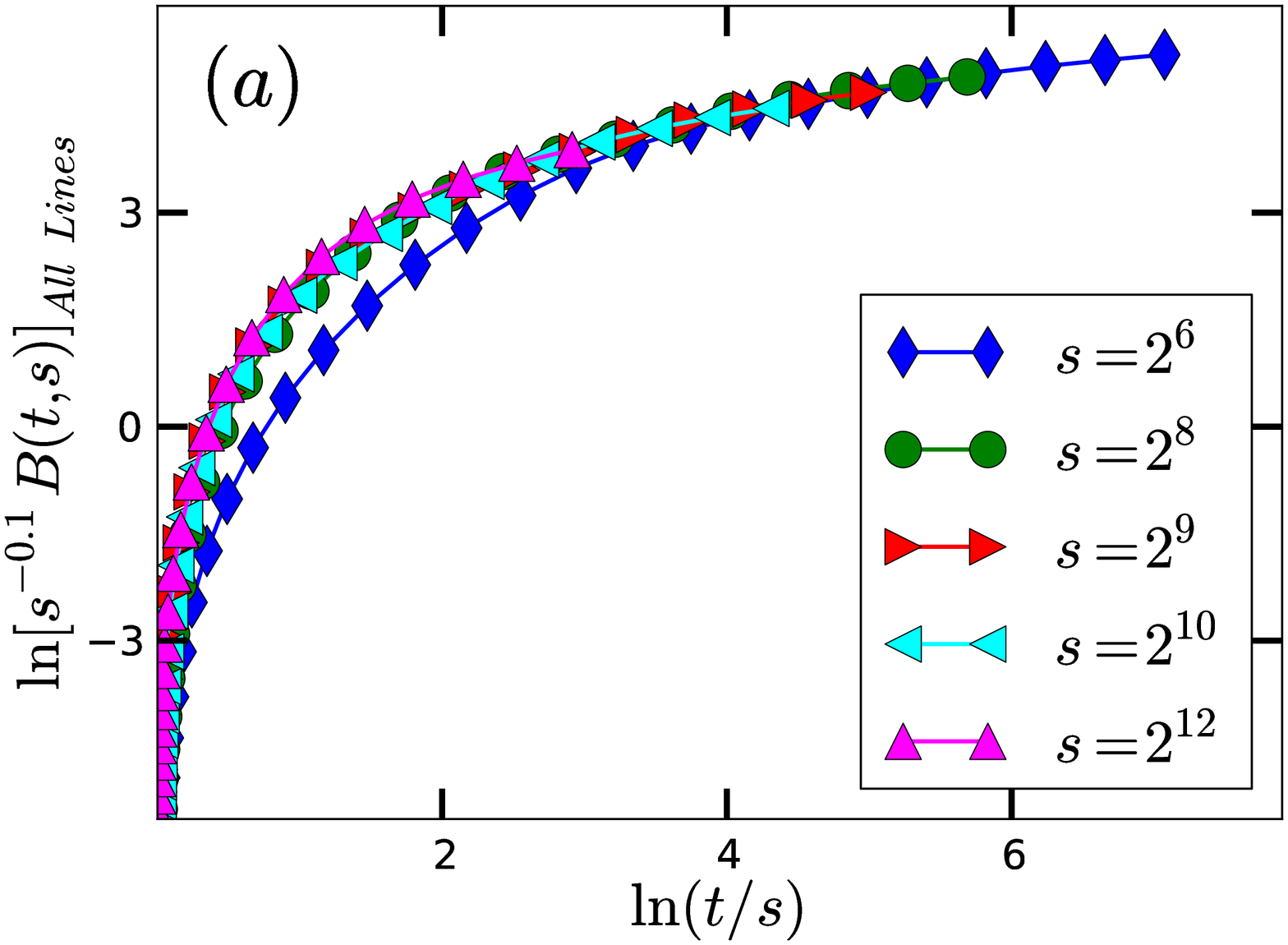}
 \label{fig:int-nodis-upquench-ball}} \\
 \subfloat{
 \includegraphics[width=0.97\columnwidth]{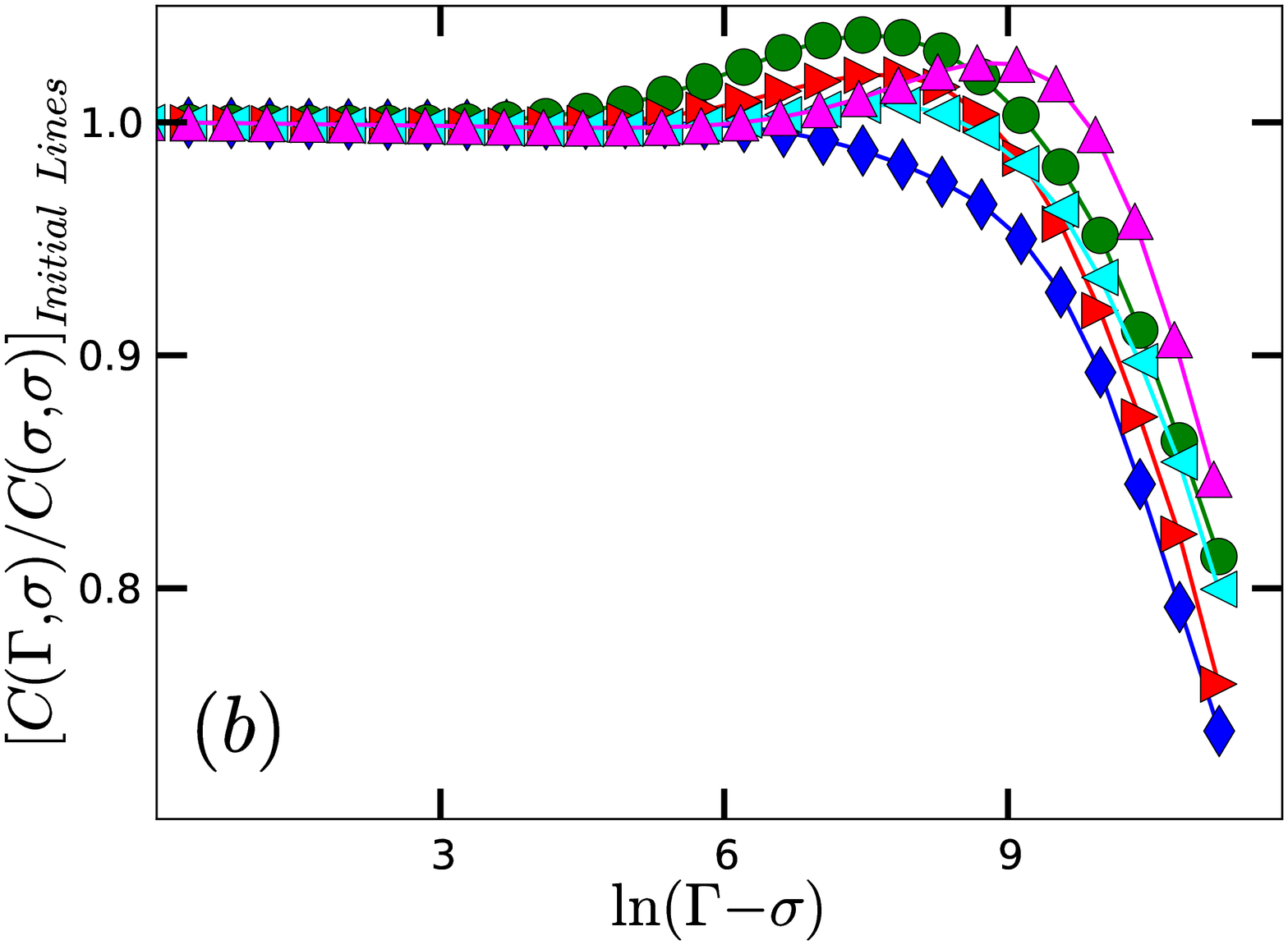}
 \label{fig:int-nodis-upquench-cin}} \\
 \subfloat{
 \includegraphics[width=0.97\columnwidth]{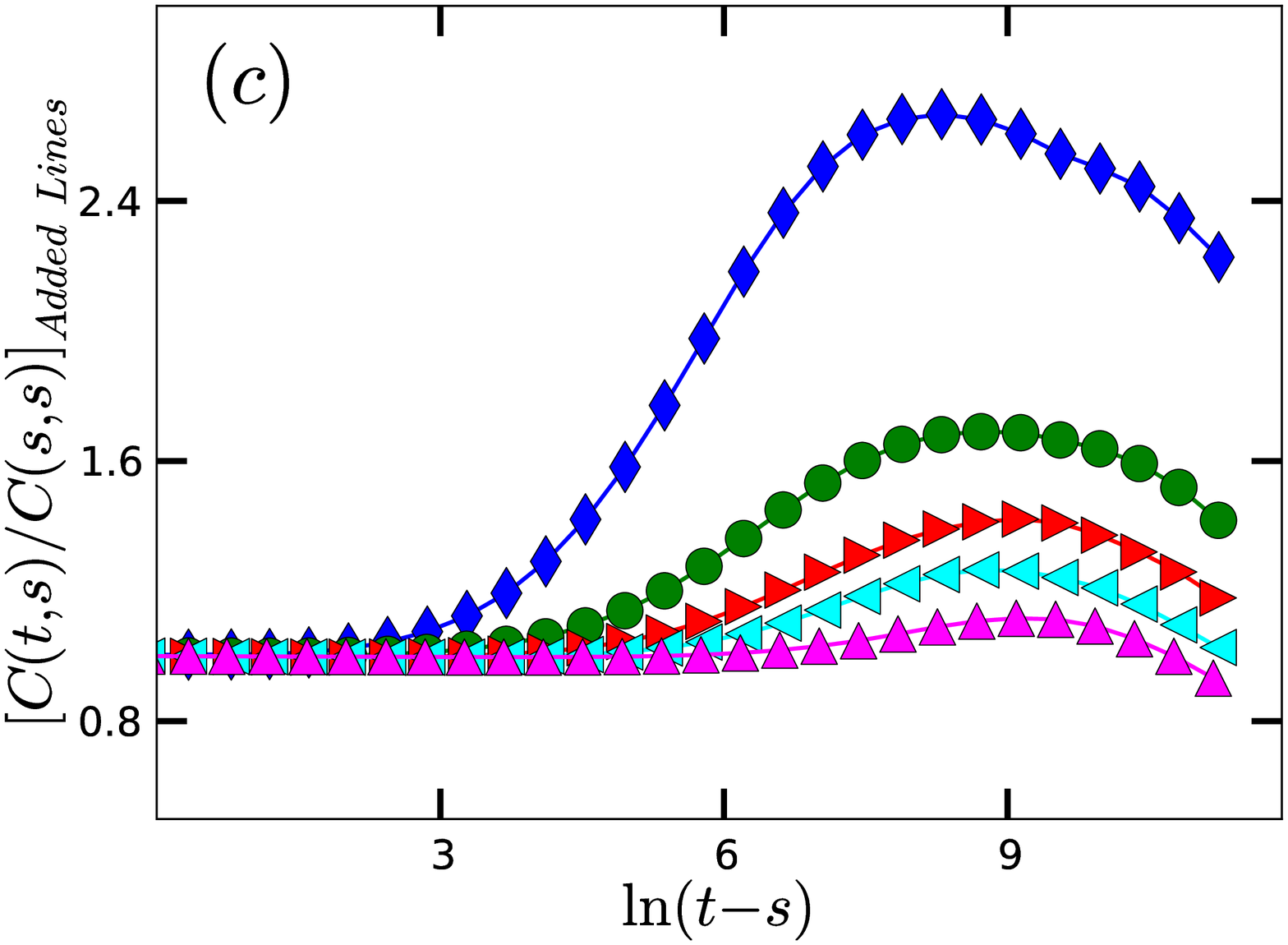}
 \label{fig:int-nodis-upquench-cad}}
 \caption{(Color online) Relaxation of (a) the mean-square displacement of all 
   the flux lines combined, the normalized height-height autocorrelation 
   function of (b) the initially present vortices, and (c) the newly added 
   lines in a system of interacting flux lines in the absence of disorder, 
   following a sudden increase in the magnetic field (data averaged over 900 
   realizations).}
\label{fig:int-nodis-upquench}
\end{figure}
We start with the system of interacting flux lines without disorder. 
At a fixed magnetic field, we observe at least approximate dynamical scaling of 
$B(\Gamma,\sigma)$ as displayed in Fig.~\ref{fig:int-nodis-noquench-b}, yet 
with a different exponent $b=1.57$ from before. 
On the other hand, the height-height autocorrelation function in 
Fig.~\ref{fig:int-nodis-noquench-c} shows a distinct two-step relaxation 
behavior, where a $\beta$ relaxation with almost no changes in 
$C(\Gamma,\sigma)$ is followed by a very slow decay. 
This is referred to as $\alpha$-$\beta$ relaxation scenario, a characteristic
feature of structural glasses~\cite{Pleimling2011,Gotze1992}. 
Furthermore, a local maximum in $C(\Gamma,\sigma)$ is observed for all waiting 
times $s$, which we attribute to the decay of metastable kink configurations. 
Such single kinks are step-like flux line structures that are stabilized by 
the screened logarithmic repulsive forces; hence they constitute 
interaction-induced features that are long-lived. 
The mean-square displacement $B(\Gamma,\sigma)$ loses the dynamical aging 
scaling property when lines are removed from the system, whereas
$C(\Gamma,\sigma)$ displays similar behavior to that at fixed magnetic field, 
but with the local maxima absent for most waiting times. 
This indicates a destabilization of single-kink structures due to the decrease
in the number of lines in the sample, which relieves the effective mutual cages 
generated by the repulsive interactions. 

On the other hand, when the magnetic field is suddenly increased and vortices
are added to the system, we observe the same time evolution in the mean-square 
displacement of the added lines, the initial lines, and all the lines 
combined. 
The data corresponding to all waiting times collapse onto a single curve with 
a long-time algebraic growth with the aging scaling exponent $b=0.1$, see 
Fig.~\ref{fig:int-nodis-upquench-ball}. 
This dynamical scaling is manifest only when the initial relaxation time $r$ 
is disregarded in our scaling formula for both the initial lines and the
collection of all lines. 
The independence of these subpopulations from the initial relaxation time 
implies that the respective lines do not retain memory of their early dynamics 
and only experience changes upon the addition of new vortices, reflecting the 
mutual interactions and consequent dynamical correlations. 

We observe a more diverse temporal evolution in the height-height 
autocorrelation functions. 
The initial lines in Fig.~\ref{fig:int-nodis-upquench-cin} show 
$\alpha$-$\beta$ relaxation for all waiting times $s$ with the presence of 
local maxima similar to those in Fig.~\ref{fig:int-nodis-noquench-c}. 
The data display a rather complicated dependence on the sequence of waiting 
times, which we cannot explain in detail but attribute to the rich and complex 
nature of the combined vortex system when these initially present and largely 
relaxed flux lines interact with newly added straight ones.
Unlike the initial line subpopulation, the added vortex kinetics presented in 
Fig.~\ref{fig:int-nodis-upquench-cad} displays a plateau followed by an 
increase, which is in turn followed by a final decrease for all waiting times.
The first increase indicates the time scale when the added lines start 
noticing the repulsive effect of the already existing vortices, whence lateral
fluctuations are enhanced. 
At later times following this initial increase, $C(t,s)$ decreases again 
indicating that the lines have reached a steady state with well-established
mutual interaction cages, and hence their fluctuations become reduced. 

In the following subsections, we investigate and analyze the effects of  
disorder on samples of interacting vortex lines.
We first address uncorrelated point-like defects and then proceed to extended 
columnar pins.

\subsection{Interacting Flux Lines with Point Disorder}
\label{sec:int-point}

\begin{figure*}
 \centering
 \subfloat{
 \includegraphics[width=0.98\columnwidth]{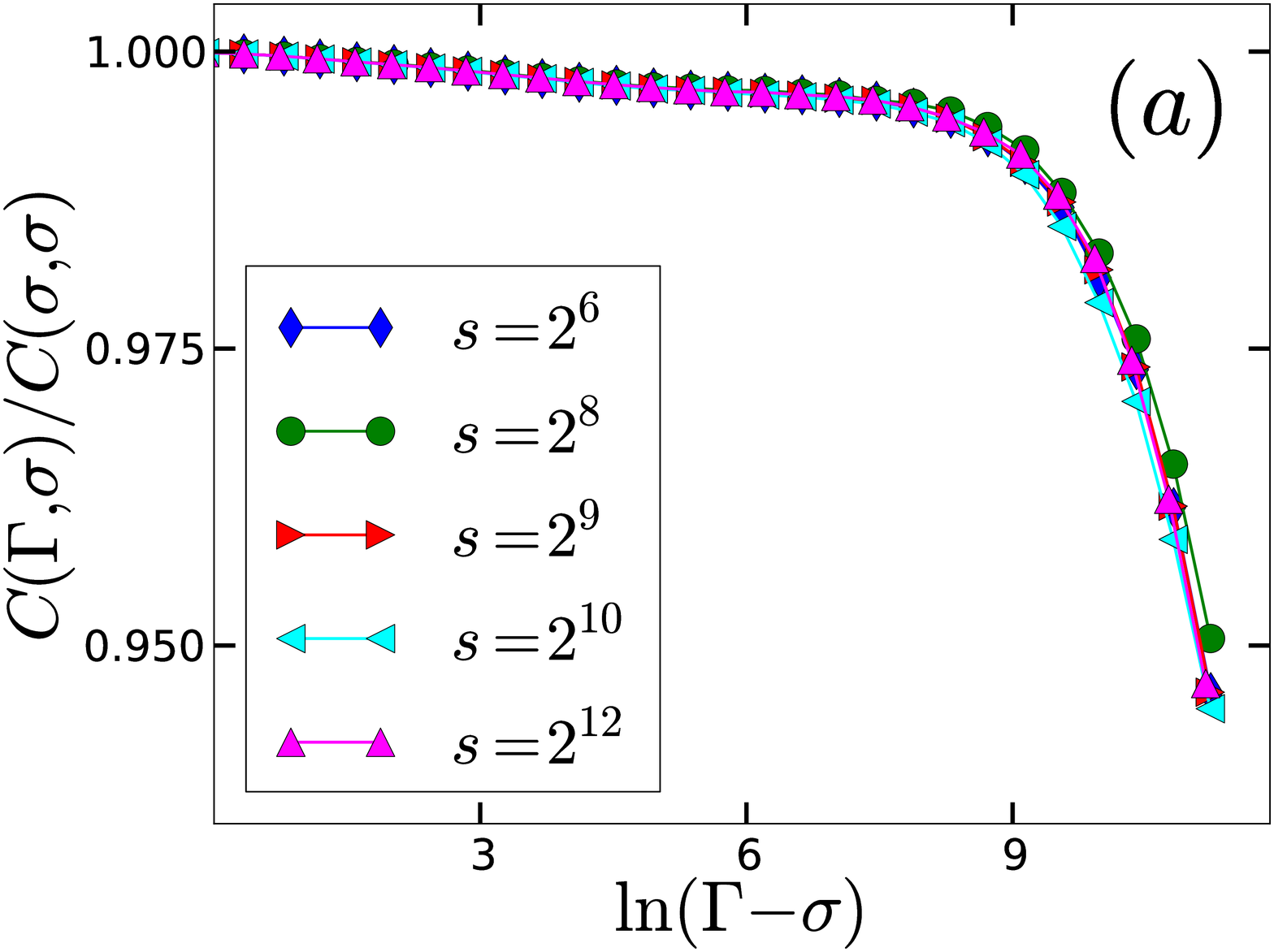}
 \label{fig:int-point-noquench-c}} \
 \subfloat{
 \includegraphics[width=0.98\columnwidth]{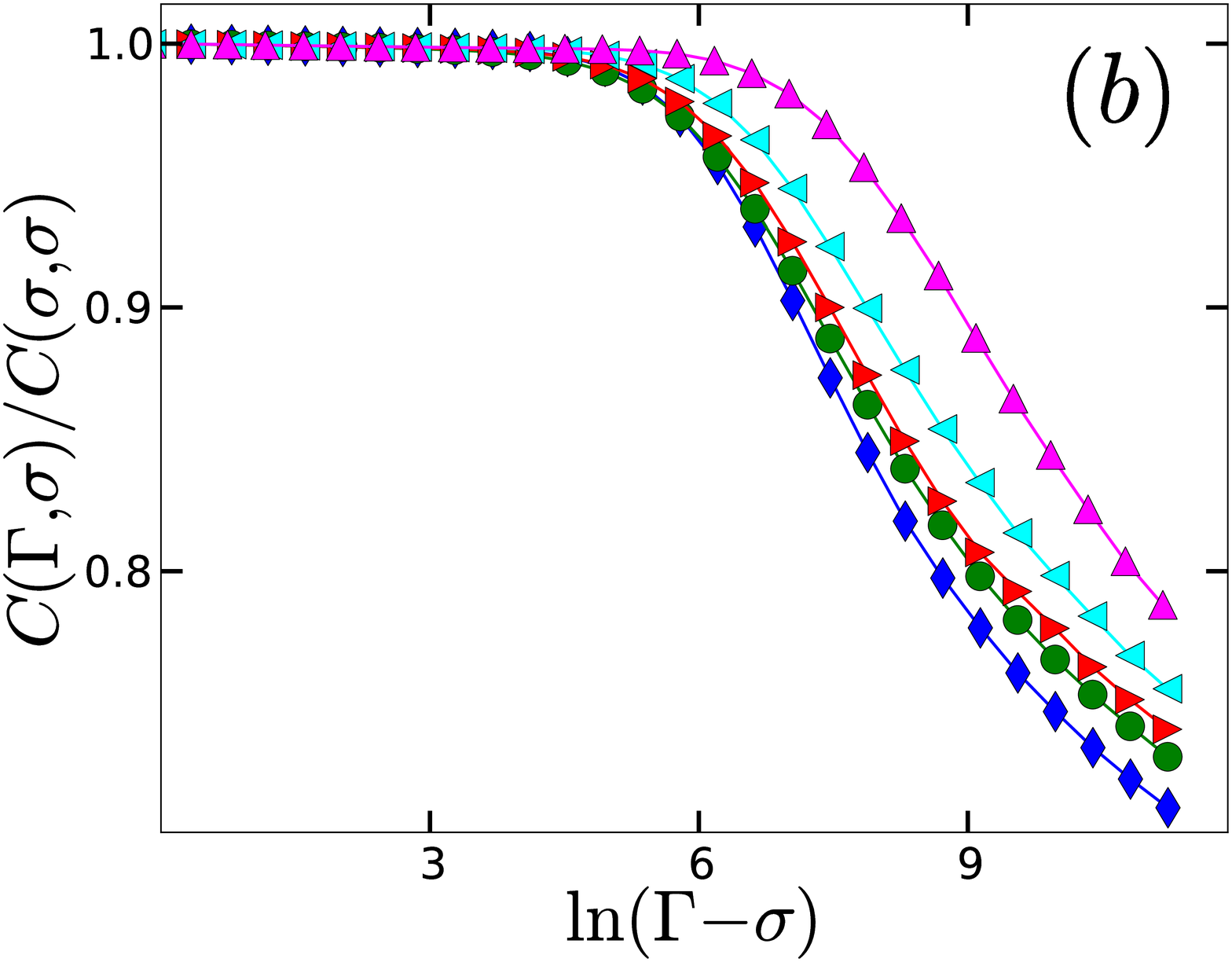}
 \label{fig:int-point-downquench-c}} \\
 \subfloat{
 \includegraphics[width=0.98\columnwidth]{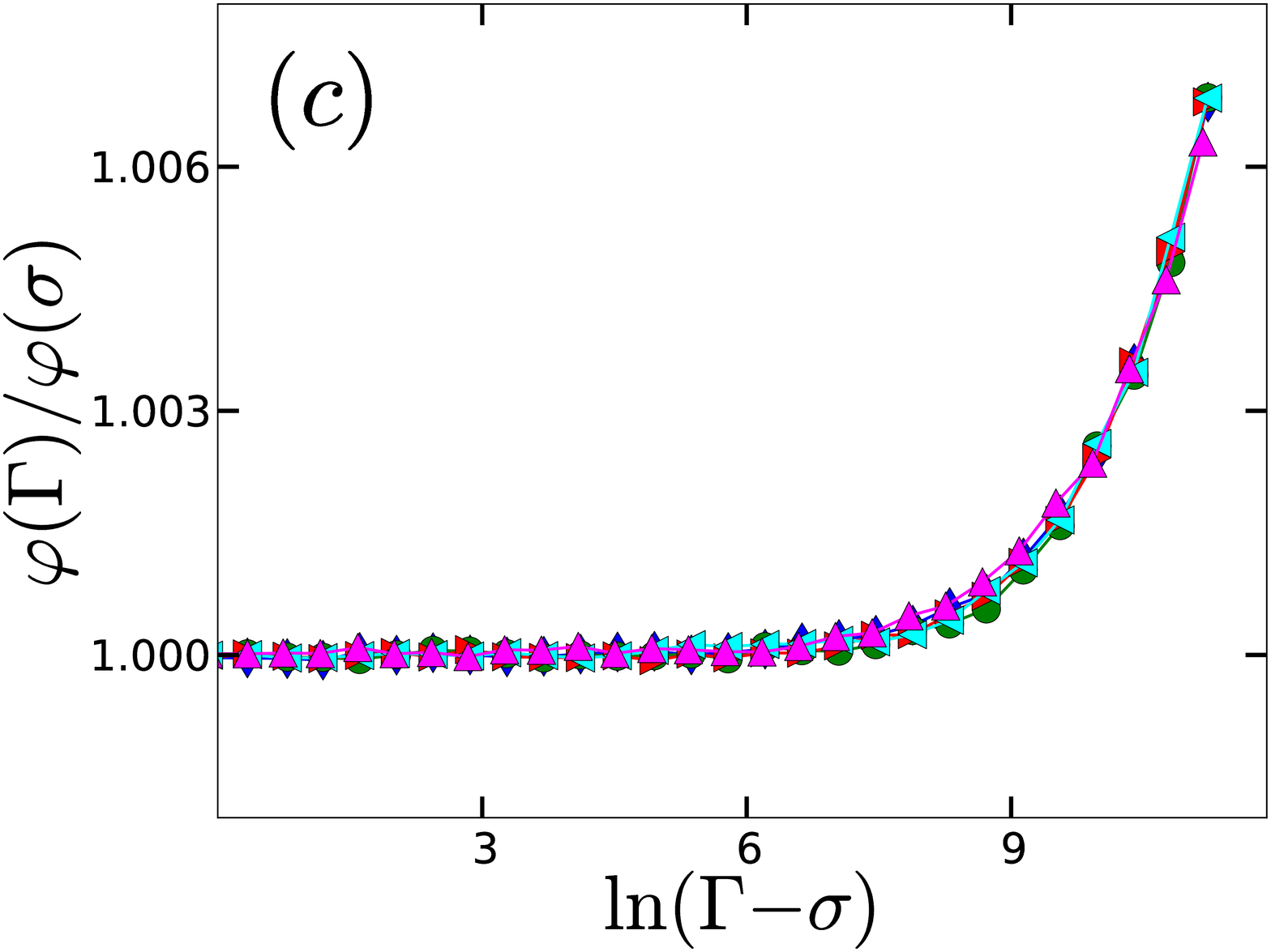}
 \label{fig:int-point-noquench-pp}} \
 \subfloat{
 \includegraphics[width=0.98\columnwidth]{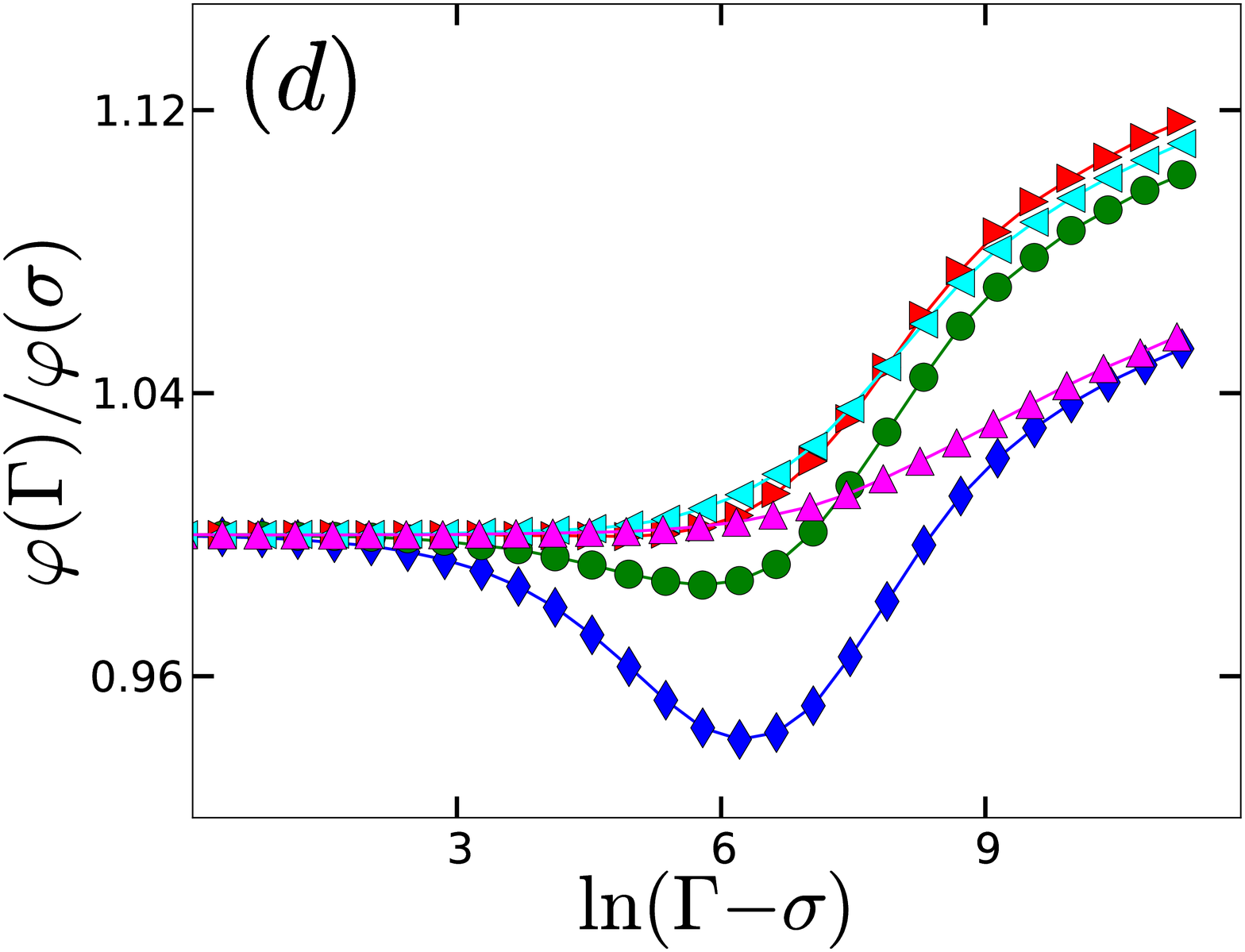}
 \label{fig:int-point-downquench-pp}}
 \caption{(Color online) Relaxation of various observables in a system of 
   interacting flux lines with point-like disorder when (a,c) the vortex 
   density remains fixed, and (b,d) after a sudden magnetic field decrease: 
   (a,b) normalized height-height autocorrelation function, (c,d) fraction of 
   pinned line elements (data averaged over 800 realizations.}
\label{fig:int-point-noquench-downquench}
\end{figure*}

Considering first the case when the magnetic field does not suddenly change, 
our sample of interacting lines in the presence of point disorder shows 
dynamical scaling for $B(\Gamma,\sigma)$ with the aging exponent $b=1.54$, 
similar to the scaling behavior displayed in 
Fig.~\ref{fig:int-nodis-noquench-b} in the disorder-free interacting vortex
system. 
This similarity suggests that the strong repulsive forces between the flux
lines dominate over the effects of point-like pinning centers, which do not
play a significant role in modifying the global relaxation dynamics of 
interacting vortices.
Even though the height-height autocorrelation function does not show dynamical
scaling, the data for the different waiting times collapse onto a single curve
indicating that the system has effectively equilibrated, see 
Fig.~\ref{fig:int-point-noquench-c}. 
The fraction of the pinned line elements $\varphi(t)$ depicted in 
Fig.~\ref{fig:int-point-noquench-pp} too shows evidence for equilibration for
all waiting times considered in our simulations.
The data display a plateau followed by an increase in the pinned line element
fraction at $\ln(\Gamma-\sigma) \approx 7$, which coincides with the time of
marked decrease in $C(\Gamma,\sigma)$, confirming that 
$\Gamma \approx 1096+\sigma$ represents a characteristic temporal scale when 
pinning to point-like defects causes significant flux line roughening. 

\begin{figure}
 \centering
 \subfloat{
 \includegraphics[width=0.97\columnwidth]{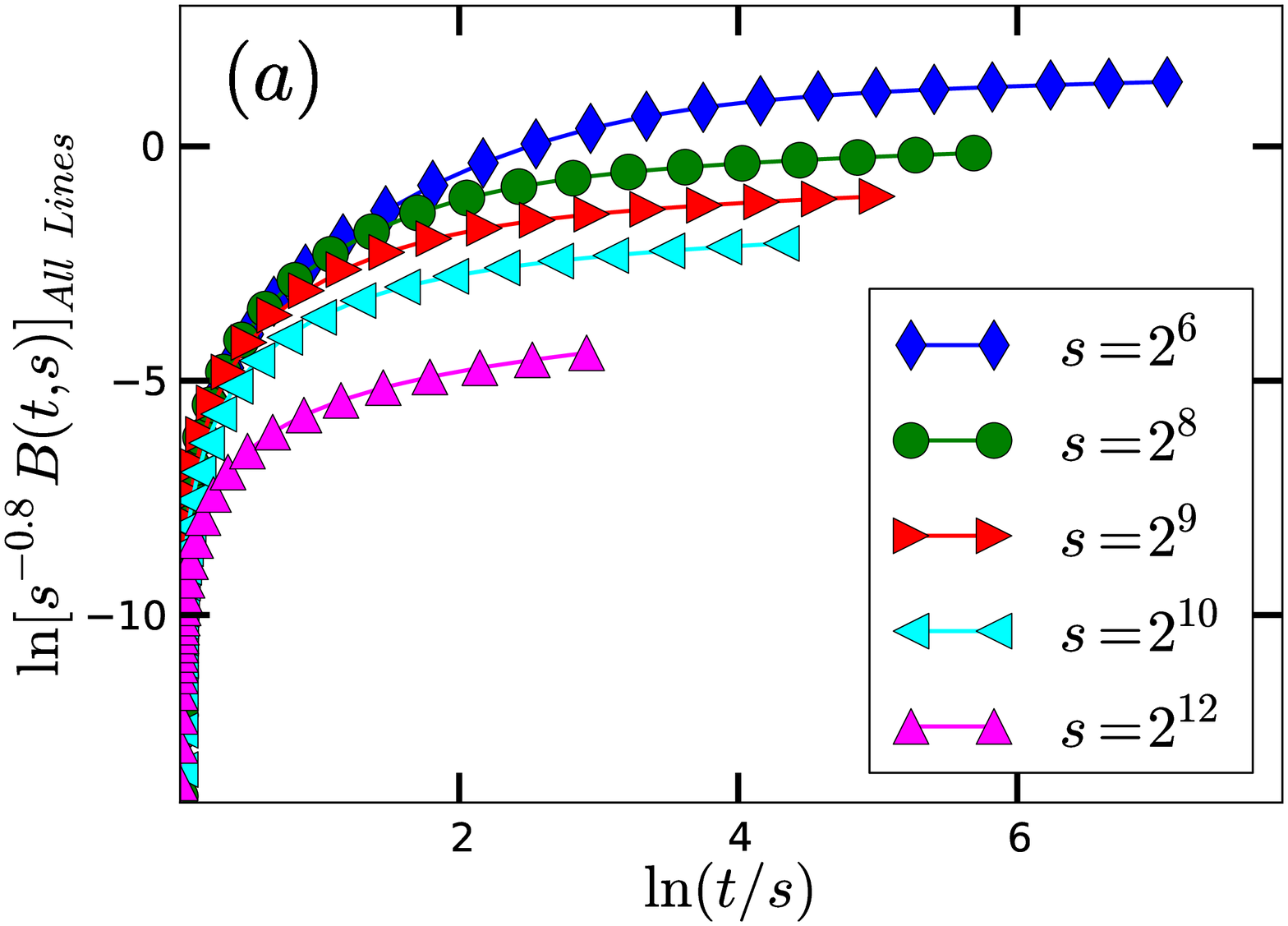}
 \label{fig:int-point-upquench-ball}} \\
 \subfloat{
 \includegraphics[width=0.97\columnwidth]{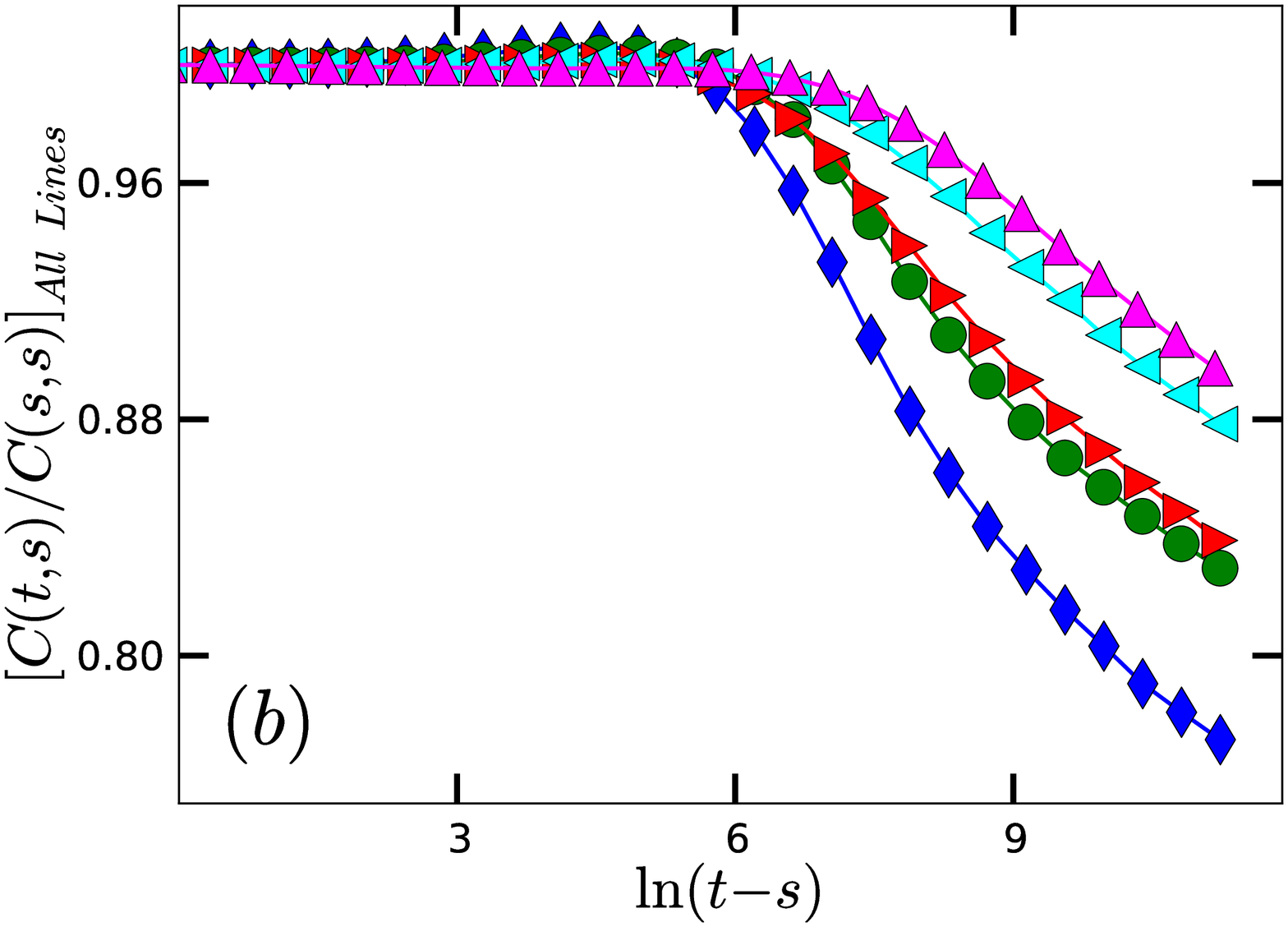}
 \label{fig:int-point-upquench-call}} \\
 \subfloat{
 \includegraphics[width=0.97\columnwidth]{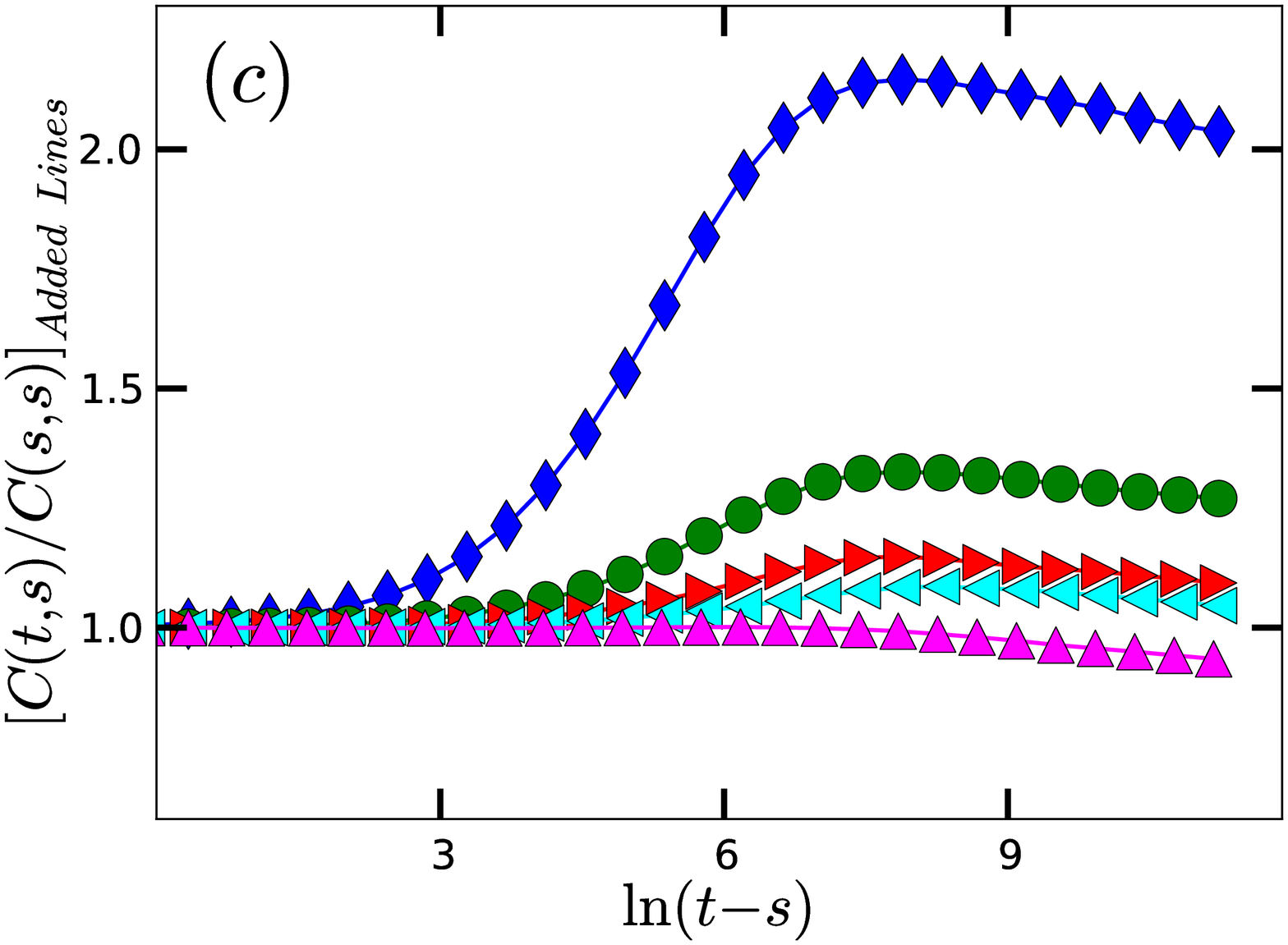}
 \label{fig:int-point-upquench-cad}}
 \caption{(Color online) Relaxation of (a) the mean-square displacement, the 
   normalized height-height autocorrelation function (b) of all the flux lines 
   combined, and (c) of the subpopulation of newly added lines in a system of 
   interacting vortices subject to point disorder, following a sudden increase
   in the magnetic field (data averaged over 800 realizations).}
\label{fig:upquench-int-nodis}
\end{figure}
When flux lines are suddenly removed from the system, $B(\Gamma,\sigma)$ of 
the remaining lines does not display dynamical scaling. 
Moreover, $C(\Gamma,\sigma)$ in Fig.~\ref{fig:int-point-downquench-c} still 
shows the $\alpha$-$\beta$ relaxation scenario for all waiting times without 
the system ever becoming fully equilibrated as in 
Fig.~\ref{fig:int-point-noquench-c}. 
The fraction of pinned line elements undergoes a more diverse temporal
evolution for the different waiting times, as can be observed in 
Fig.~\ref{fig:int-point-downquench-pp}. 
For short waiting times $s \leq 2^8$, a minimum in $\varphi$ occurs at 
$\ln(\Gamma-\sigma) \approx 6.2$.  
The initial decrease indicates that the sudden release of caging due to 
removed lines leads to depinning of some remaining vortices from point defects 
where the line elements were only held in place because of their interactions
with the removed lines. 
This enables the remaining flux lines to explore a farther range of their 
surroundings, ultimately becoming bound to new and more  point pins to 
eventually reach a steady-state fraction $\varphi$ that is a little larger 
than the initial one. 
For intermediate-to-larger waiting times $s > 2^9$, this minimum in the pinned
line element fraction is not visible and the data looks similar to the 
situation with fixed vortex density, see Fig.~\ref{fig:int-point-noquench-pp}; 
the only difference being that the quenched system has clearly not 
equilibrated yet.

Let us now explore the effects of a sudden increase in the magnetic field, 
where new straight flux lines are added to the a previously relaxed system of 
interacting vortices in the presence of point disorder, and analyze the 
resulting dynamics. 
Dynamical scaling of the mean-square displacement is lost for the newly added 
lines, the initially present vortices, and all the flux lines combined. 
However, the mean-square displacement as function of time looks similar among 
all these different subpopulations; an example is shown in 
Fig.~\ref{fig:int-point-upquench-ball}. 
Note that our choice of a putative aging scaling exponent $b\approx 0.8$ in 
Fig.~\ref{fig:int-point-upquench-ball} is merely to demonstrate the global 
dynamical evolution of $B(t,s)$ for the different vortex subpopulations. 

On the other hand, the height-height autocorrelation function shows similar 
results for the initial lines and the combination of all vortices, as shown in 
Fig.~\ref{fig:int-point-upquench-call}, where $\alpha$-$\beta$ relaxation is 
evident as a characteristic feature in these flux line subpopulations that 
have already relaxed for a long time period, but are still not equilibrated.  
This lack of equilibration becomes manifest in the height-height 
autocorrelation function for just the subpopulation of added lines in 
Fig.~\ref{fig:int-point-upquench-cad}. 
For these newly introduced vortices, only very long waiting times 
$s \geq 2^{12}$ display $\alpha$-$\beta$ relaxation kinetics comparable to 
the situation at fixed magnetic field. 
However, data corresponding to $s<2^{12}$ show an initial increase in $C(t,s)$ 
followed by a slow decrease. 
When new flux lines are inserted into the sample, the originally present 
vortices repel them and thus facilitate their pinning to the uncorrelated 
point disorder, which enhances transverse line fluctuations. 
At $\ln(t-s) \approx 7$, the added lines' height autocorrelation function 
decreases because the lines have found a (near) optimal pinning configuration. 
The fraction of pinned line elements behaves analogously to that in the case 
of a sudden vortex density decrease, Fig.~\ref{fig:int-point-downquench-pp}, 
with the mere difference that the minimum is less pronounced for a sudden 
magnetic field up-quench.
Yet in this case, the sudden presence of additional caging potentials (rather
than their release) and subsequent vortex rearrangement leads to the observed
behavior of the pinned line element fraction.

\subsection{Interacting Vortices with Columnar Defects}
\label{sec:int-col}

Finally, we analyze a sample of interacting flux lines in the presence of 
columnar defects, and compare the relaxation kinetics in this system with the 
preceding investigation of uncorrelated point disorder.

\begin{figure}
  \centering
  \includegraphics[width=0.98\columnwidth]{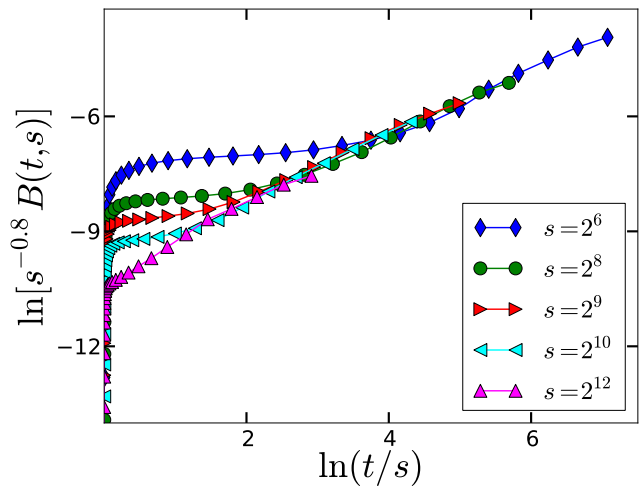}
  \caption{(Color online) Relaxation of the mean-square displacement in a 
    system of interacting flux lines with columnar defects, when the vortex 
    density remains fixed (data averaged over 1500 realizations).}
  \label{fig:int-col-noquench-b}
\end{figure}

We first study the case when the magnetic field stays constant. 
We find that $B(t,s)$ shows dynamical scaling at long times with the aging 
scaling exponent $b=0.8$, see Fig.~\ref{fig:int-col-noquench-b}. 
In contrast, no dynamical scaling is obtained for $C(\Gamma,\sigma)$ which
displays a time evolution similar to that in 
Fig.~\ref{fig:int-point-noquench-c} for point defects. 
The fraction of pinned line elements in the presence of columnar defects turns
out to not significantly differ from that in the presence of point disorder at 
fixed vortex density. 

When flux lines are removed from the sample, dynamical scaling is not observed 
for $B(\Gamma,\sigma)$, even at long times. 
Furthermore, time translation invariance in the normalized height-height 
autocorrelation function indicates an essentially equilibrated system that 
undergoes $\alpha$-$\beta$ relaxation. 
This behavior is accompanied by an increasing fraction of pinned line elements
confirming that the remaining flux lines undergo similar relaxation dynamics 
as in the system at fixed magnetic field, namely a continuous straightening 
and hence reduced transverse fluctuations as the vortices find nearby columnar 
pinning centers.

\begin{figure}
 \centering
 \subfloat{
 \includegraphics[width=0.97\columnwidth]{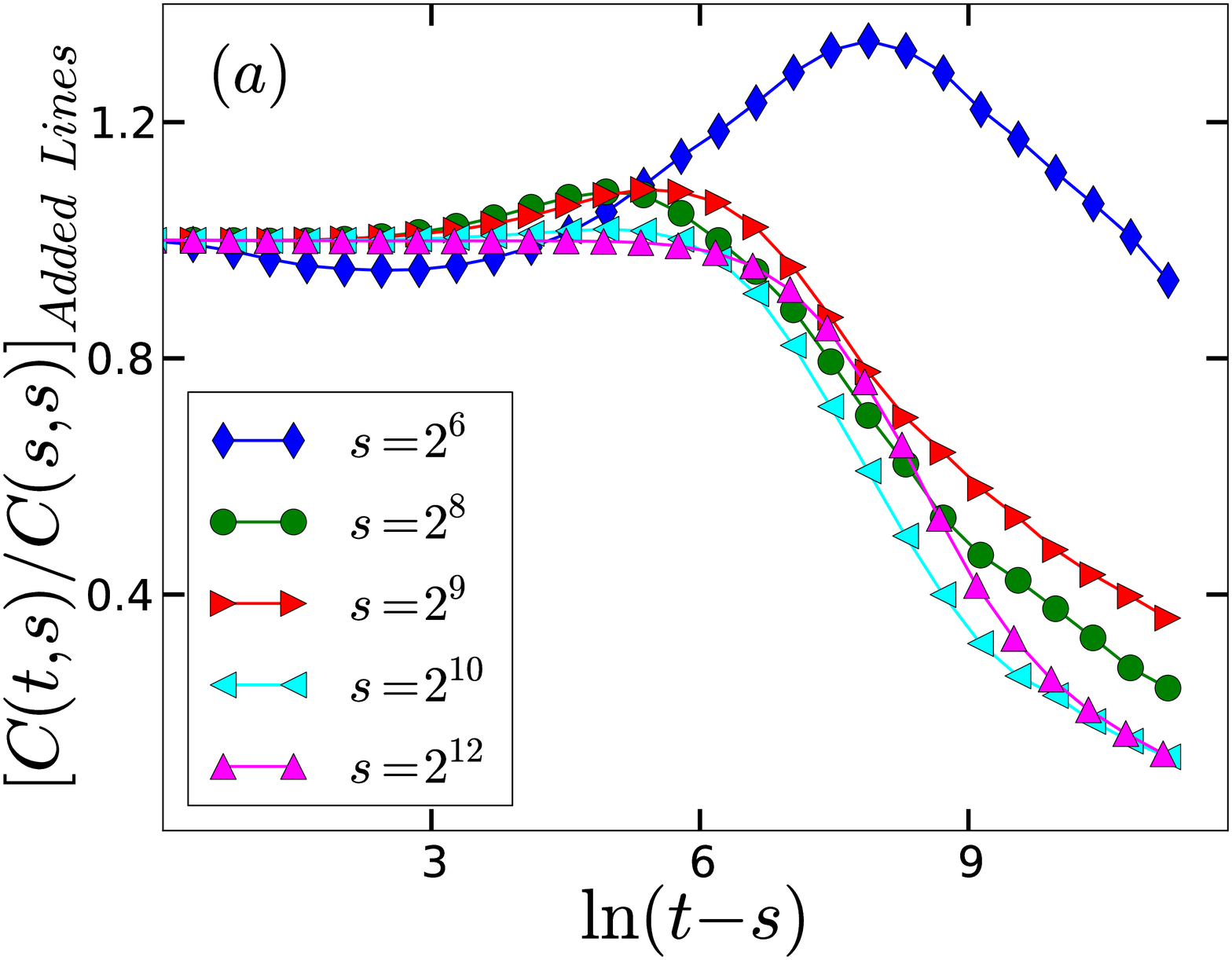}
 \label{fig:int-col-upquench-cad}} \\
 \subfloat{
 \includegraphics[width=0.97\columnwidth]{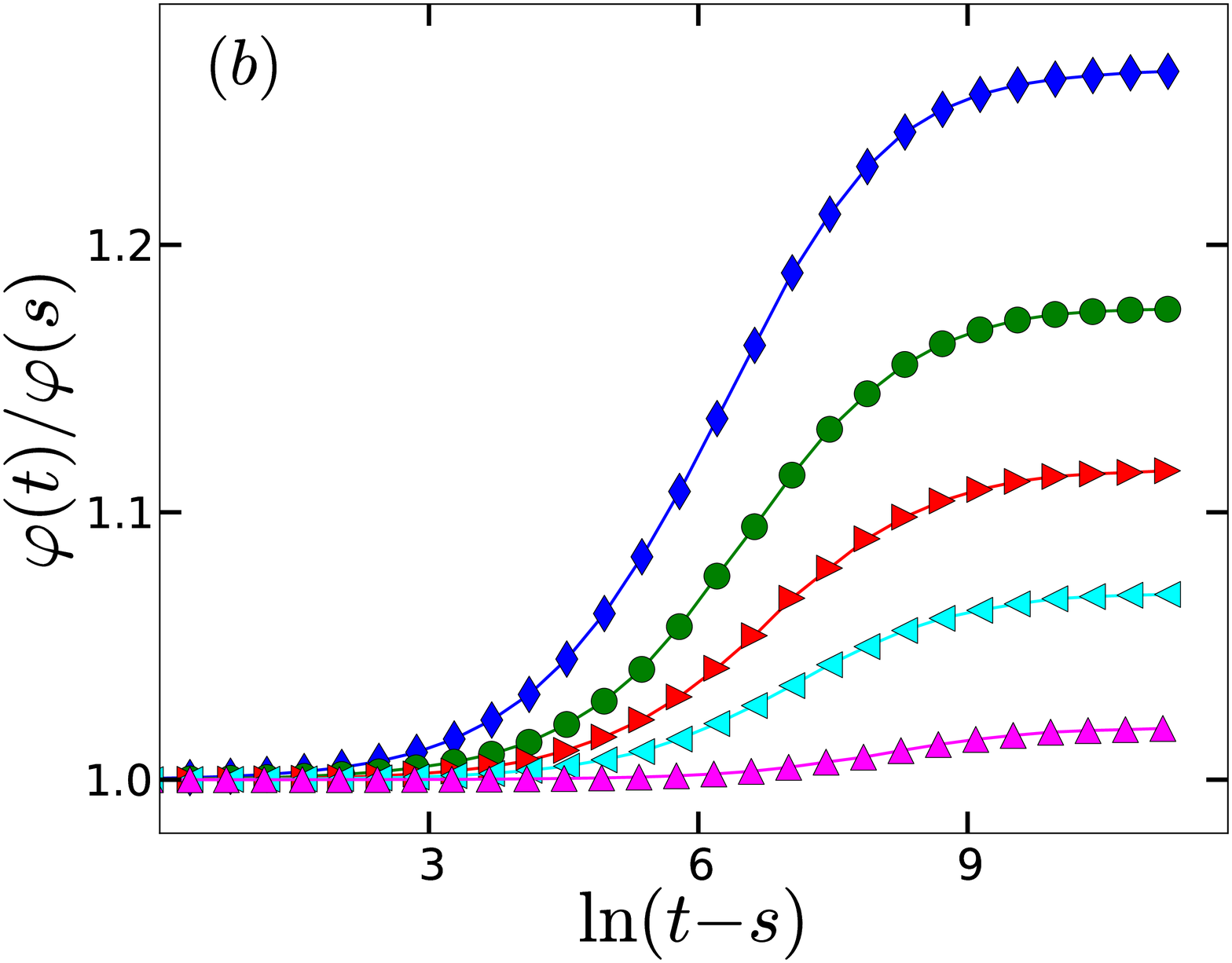}
 \label{fig:int-col-upquench-pp}} 
 \caption{(Color online) Relaxation of (a) the normalized height-height 
   autocorrelation function of the newly added flux lines, and (b) the fraction 
   of pinned line elements in a system of interacting vortices with columnar 
   defects, following a sudden increase in the magnetic field (data averaged 
   over 800 realizations).}
\label{fig:int-col-upquench}
\end{figure}
Quite different features are observed when the magne\-tic field is increased 
instantaneously: 
The temporal growth of the mean-square displacement $B(t,s)$ of the newly 
introduced lines looks similar to samples with point-like disorder as shown in 
Fig.~\ref{fig:int-point-upquench-ball}. 
The similarity in the mean-square displacement in the presence of both point
and columnar pinning sites underscores our earlier observation that 
introducing disorder into the interacting vortex system has comparatively 
little effect on the global dynamics which is highly dominated by the strong
repulsive forces between the flux lines. 
Analogous to the case when point disorder is present, all the different vortex
subpopulations show similar results for the mean-square displacement. 

On the other hand, the added vortices and the initially present flux lines are
characterized by different height autocorrelations. 
The initial vortices as well as the combination of all lines display 
$\alpha$-$\beta$ relaxation for all waiting times $s$. 
The appearance of local maxima for some long waiting times indicate the decay 
of long-lived vortex double kinks in the system. 
However, the subpopulation of added lines displays different features in their
height autocorrelations $C(t,s)$ for the different waiting times, see 
Fig.~\ref{fig:int-col-upquench-cad}. 
For short waiting times, {\em e.g.} $s=2^6$, an initial decrease is followed 
by a pronounced maximum in $C(t,s)$. 
The initial decrease is set by the time scale when the added flux lines begin 
to notice the pinning centers, causing their local transverse fluctuations to 
decrease. 
The following increase indicates the characteristic time range when the 
additional vortices start feeling the repulsive interactions with the 
initially present flux lines and with one another; this produces enhanced 
fluctuations about their mean lateral positions.  
This increase continues until pinning effects start to again dominate the 
relaxation dynamics, rendering the lines straighter and thus reducing vortex
line fluctuations. 
For longer waiting times, $\alpha$-$\beta$ relaxation becomes more prominent, 
with small local maxima present for waiting times $s=2^8$ and $s=2^9$. 
Moreover, the fraction of pinned line elements in 
Fig.~\ref{fig:int-col-upquench-pp} shows an increase followed by asymptotic
saturation for all waiting times tested in our simulations.

\section{Conclusion and Outlook}
\label{sec:conclusion}

In this paper, we have explored the effects of incorporating experimentally
motivated initial conditions into the study of the non-equilibrium relaxation 
dynamics of magnetic flux lines in type-II superconductors. 
Random initial conditions were considered in our earlier work in 
Refs.~\cite{Pleimling2011,Dobramysl2013}. 
In the present study, we focused on the effects of sudden changes in 
temperature and magnetic field on the ensuing vortex relaxation properties. 
We investigated the differences between the relaxation kinetics of flux lines
in the presence of uncorrelated point-like disorder and columnar defects. 
To this end, we modeled the magnetic vortices as directed elastic lines and 
employed a Langevin molecular dynamics algorithm to simulate the vortex 
kinetics. 

We studied the relaxation of a system of interacting flux lines in the 
presence of disorder at fixed temperatures to confirm that higher temperatures
induce enhanced thermal fluctuations that counteract the straightening of the
vortex lines caused by their pinning to columnar defects, while in contrast
these same fluctuations help lines to roughen when binding to uncorrelated 
point-like disorder. 
Furthermore, we investigated the relaxation processes following sudden changes
in the ambient temperature. 
Systems subject to the two types of disorder show very distinct relaxation 
dynamics when the temperature is instantaneously increased due to the 
different (de-)pinning behavior from these defects. 
Some flux line elements become depinned from the uncorrelated point defects 
following a sudden increase in the sample's temperature confirming that this 
type of disorder enhances the thermal wandering of flux lines. 
However, samples with columnar defects display a non-monotonic behavior in 
spatial fluctuations due to the creation and subsequent decay of vortex
double-kinks.
These pronounced differences between systems with point-like and columnar 
defects are not present when the temperature is suddenly reduced, since this 
decrease simultaneously binds more line elements to the pinning centers, 
regardless of the type of disorder present in the sample.   

We carefully investigated the effects of sudden changes in the magnetic field,
{\em i.e.}, vortex density, on the relaxation processes of flux lines, and 
performed a systematic analysis to disentangle the contributions of the 
repulsive vortex-vortex interactions and pinning to the distinct point-like 
and columnar defects. 
We compared the cases when the magnetic field instantaneously increases or 
decreases to the situation when it stays fixed. 
We start with perfectly straight and randomly placed vortex lines and let the 
system relax for an extended relaxation time period, after which we maintain a
fixed number of lines representing a constant magnetic field. 
Alternatively, we remove randomly selected lines from the initially relaxed 
vortices in the sample, or add perfectly straight new flux lines at random 
positions into the system of relaxed lines, to analyze the effects of a sudden 
decrease or increase in the magnetic field, respectively. 

We validated our numerical studies by showing that our results for free, 
non-interacting vortex lines agree with the analytical predictions from the 
Edwards-Wilkinson interface growth model. 
We then introduced point-like and columnar pinning centers to a system of 
non-interacting flux lines. 
We found that the relaxation features in samples with columnar disorder are
quite similar to the case when no disorder is present, especially in the 
behavior of global observables, as {\em e.g.}, visible in the dynamical
scaling of the vortex mean-square displacement. 
This similarity indicates that the dynamics of vortex lines in the presence 
of columnar disorder is strongly dominated by the still unattached lines in
the system. 
In the three studied cases without vortex interactions, sudden removal of flux
lines results in the remaining lines showing similar relaxation behavior to 
the situation with a fixed vortex density, as expected since caging effects 
due to the repulsive vortex-vortex interactions are absent. 
On the other hand, when additional lines are introduced to these systems, the
initially existing lines and the added lines show quite distinct features in 
their relaxation dynamics. 

Thereafter, we added the repulsive screened logarithmic interactions into our 
consideration of systems without disorder, with uncorrelated point-like, or 
with extended columnar defects. 
Sudden decreases in the magnetic field result in the remaining lines 
displaying relaxation properties that differ from the situation when the
magnetic field stays fixed, which confirms that caging effects due to 
vortex-vortex interactions constitute a prominent factor that highly affects 
the results of sudden changes in the magnetic field.
Collective interaction effects are markedly enhanced when accompanied with 
disorder following sudden increases in the magnetic field, where the 
relaxation properties of the added lines clearly depend on the type of pinning
centers. 
Point-like disorder enhances transverse spatial vortex line fluctuations by 
binding each to many pinning centers at once in the vortex glass phase, while 
columnar defects pin entire flux lines and straighten them in the distinct
Bose glass phase. 
It is worth noting that when comparing the novel results in this paper to our 
earlier study in Ref.~\cite{Dobramysl2013} with random initial conditions, we 
discovered improved dynamical scaling in the vortex mean-square displacement 
in all situations considered here.
Consequently, the currently utilized more realistic initial conditions help 
the system to equilibrate faster and thus acquire universal features.

The goal of this study is to methodically work towards observables and 
protocols that can be realized in experiments, while simultaneously providing 
detailed theoretical insight into the non-equilibrium relaxation kinetics of 
disordered vortex matter. 
Due to the complex nature of interacting vortex lines in the presence of 
disorder, we need to carefully and systematically analyze the contributions of 
the different energetic contributions to fully understand this intriguingly 
complicated dynamical system, as well as assess the relevance of different 
initial configurations. 
Hence the implementation of temperature and magnetic field quenches in our
numerical simulations constitutes a crucially important advance that should 
hopefully provide valuable insights for future experiments.
Indeed, temperature or magnetic field quenches and the subsequent measurement 
of one-time observables and two-time correlation functions could be utilized 
to formulate experimental methods to characterize material samples by means of
the observed fluctuation spectrum. 
In future work, we plan to extend our study to include the effects of driving 
currents on the transient properties of driven flux line systems, where 
relaxation towards non-equilibrium steady states will ensue in contrast to the
presently analyzed relaxation towards thermal equilibrium. 
A more detailed understanding of relaxation processes and the origin of 
deviation from scaling requires identifying the characteristic length scales 
in the system and their temporal growth, which we hope to pursue through the 
analysis of space-time observables.

\section*{Acknowledgments}
\label{sec:acknowledgements}
This research is supported by the U.S. Department of Energy, Office of Basic 
Energy Sciences, Division of Materials Sciences and Engineering under Award 
DE-FG02-09ER46613.

\end{document}